\documentclass[10pt]{iopart}

\usepackage[english]{babel}
\usepackage{hyperref}
\usepackage[]{graphicx}
\usepackage{amsmath}
\usepackage{braket}

\begin{document}

\title{Quantum model for Impulsive Stimulated Raman Scattering}


\author{Filippo Glerean$^1$, Stefano Marcantoni$^{1,2}$, Giorgia Sparapassi$^1$, Andrea Blason$^1$, Martina Esposito$^3$, Fabio Benatti$^{1,2}$ and Daniele Fausti$^{1,4}$}
\address{$^1$ Dipartimento di Fisica, Università degli studi di Trieste, Via Valerio 2 Trieste 34127, Italy}
\address{$^2$ Istituto Nazionale di Fisica Nucleare, Sezione di Trieste, Trieste 34014, Italy}
\address{$^3$ Clarendon Laboratory, Department of Physics, University of Oxford, Oxford OX1 3PU, United Kingdom}
\address{$^4$ Sincrotrone Trieste S.C.p.A., Basovizza 34127, Italy}

\eads{\mailto{benatti@ts.infn.it}, \mailto{daniele.fausti@elettra.eu}}


\date{\today}

\begin{abstract}    
The interaction between ultrashort light pulses and non-absorbing materials is dominated by Impulsive Stimulated Raman Scattering (ISRS). The description of ISRS in the context of pump\&probe experiments is based on effective classical models describing the interaction between the phonon and pulsed electromagnetic fields. Here we report a theoretical description of ISRS where we do not make any semi-classical approximation and we treat both photonic and phononic degrees of freedom at the quantum level. 
The results of the quantum model are compared with semiclassical results and validated by means of spectrally resolved pump\&probe measurements on $\alpha$-quartz.
\end{abstract}


\maketitle

\ioptwocol


\section{Introduction}
The excitation and measurement of coherent lattice (or molecular) vibrations in time domain experiments rely on the possibility of using ultrashort optical pulses in pairs, one as a pump and a second as a probe. The pump should be capable of injecting energy into the phonon modes on time scales shorter than the inverse of the phonon frequency, while the probe should be short enough to measure the time evolution of this state with time. In this limit, photoexcitation produces coherent vibrational states whose dissipative dynamics can be directly accessed by pump\&probe experiments \cite{bib:Hase, bib:Hase2, bib:Hase3, bib:Fahy, bib:Merlin3, bib:Misochko, bib:Misochko2, bib:Misochko3}. 

The processes for transferring energy from the optical pulse to the phonons depends strongly on the nature of the material. In absorbing systems, the whole light-matter interaction processes should be described taking into account that the dissipative dynamics affecting the photo-excited electrons, which mediate the energy transfer from the light pulse to lattice excitations, may play a crucial role \cite{bib:Zeiger, bib:Merlin1, bib:Merlin2, bib:Fausti, bib:Papalazarou, bib:Randi, bib:revSpectro}. The situation is simpler in ``transparent materials'', i.e. in materials where there is no dipole allowed electronic transitions available in the frequency range of the ultrashort pulses. In this limit, the interaction between the latter and the vibrational modes is a coherent process, where dissipative electron dynamics can be neglected and the whole process can be described effectively as a direct coupling between the ultrashort pulses and the phonon modes.

In this limit the interaction is dominated by processes dubbed Impulsive Stimulated Raman Scattering (ISRS) \cite{bib:ISRSNelson, bib:Nelson2}. ISRS takes place whenever a sufficiently short laser pulse (i.e. characterized by a multimode frequency spectrum larger than the phonon frequency) propagates through a Raman-active medium. In this limit, components of the electric field at different frequency can interact through ISRS if the difference between their photon energy matches the energy of a phonon mode. More formally, ISRS can be described as a coherent process mixing three frequencies\footnote{The process occurs simultaneously between all pairs of frequencies and therefore the resulting electric field at frequency $\omega$ is influenced by both the electric field components at $\omega \pm \Omega$.} where the stimulated annihilation (creation) of a photon of frequency $\omega$ occurs simultaneously to the creation (annihilation) of one at frequency $\omega \pm \Omega$. The overall process can create (Stokes) or annihilate (anti-Stokes) an excitation in the system and thereby result in lattice excitations. Note that in literature the term ISRS is often used to describe the whole pump and probe measurement process, while here we use it solely to indicate the physical process describing photon-phonon interaction. We will apply ISRS separately to describe how coherent lattice vibrations can be produced (pump) and measured (probe) by ISRS. 

In this paper we study the leading processes occurring in a pump\&probe experiment in transparent materials. From the interaction energy, given as a scalar product of the electric and the polarization fields, we construct the quantum Raman Hamiltonian which rules the bulk dynamics. In addition to this, we consider a modulation of the refractive index of the material that we refer to as Linear Refractive Modulation (LRM). The achronim highlights the fact that LRM consists of a "Linear"\footnote{We note that the overall process is describing non-linear responses in the susceptibility, but the term "linear" is used to clarify that no frequency mixing of the probe spectral components occurs.} interaction of the probe with a material whose Refractive index is Modulated in time by the evolving atomic position. We stress that while ISRS is a non-linear process coupling different spectral components of the pulses, LRM does not mix different probe frequencies. 

Our approach provides a formalism to describe the fundamental differences between the ISRS and the LRM. LRM amounts to a modulation of the refractive index induced by the instantaneous position of the atoms (phonon position operator, $q$) which does not couple different photonic mode operators ($a_j$), but induces a change in transmittivity or polarization which is uniform with respect to the spectral components. Conversely, ISRS is a nonlinear process that produces a shift of the spectral weights relative to different photonic modes in the probe pulses \cite{bib:Nakamura} and follows the phonon momentum operator, $p$. ISRS and LRM give rise to a time-oscillation of the response with the same frequency but shifted in phase and, more importantly, with different spectral content. The two processes are often observed simultaneously and can result in composite responses. We validate the proposed quantum model by comparing the results with those obtained by classical calculations \cite{bib:Merlin0, bib:RevNelson} and with the outcomes of pump\&probe experiments in $\alpha$-quartz providing time and spectral resolution of the probe pulses. Further, we show that it is possible to disentangle experimentally LRM and ISRS effects in pump\&probe experiments by selecting a proper combination of polarizations exploiting the symmetry of the crystal \cite{bib:Rundquist}. 

The paper is structured as follows. In section \ref{sec: model} we describe the quantum model for light-matter interaction. In particular, we distinguish between the peculiar characteristics of LRM and ISRS not always recognized in the literature \cite{bib:RevNelson, bib:Righini}. In section \ref{sec: pumpprobe} we apply the general model already discussed in the context of pump\&probe experiments, highlighting similarities and differences between the pumping and the probing processes, mainly due to the different vibrational target states before the photon-phonon interaction. In section \ref{sec: quartz} the results are validated by means of spectrally resolved pump\&probe experiments on $\alpha$-quartz, where combination between pump and probe polarizations allow for the experimentally accessible distinction of ISRS and LRM processes. Finally, we conclude with some remarks and new perspectives offered by the fully quantum treatment of time-domain experiments \cite{bib:Nori1, bib:Nori2, bib:Martina, bib:Misochko, bib:Hussain}. 

\section{Light-phonon interaction}
\label{sec: model}

A dielectric medium is polarized by an electromagnetic wave propagating through it. 
The components of the polarization field $\vec{P}$ are expressed in terms of the impinging electric field $\vec{E}$ and the material susceptibility tensor $\boldsymbol{\chi}$:
\begin{equation}
P_\lambda = \epsilon_0 \sum_{\lambda'}\chi_{\lambda\lambda’}E_{\lambda’},
\end{equation}
where $\epsilon_0$ is the electric permittivity of the vacuum.
One of the fundamental ingredients of the whole discussion is the susceptibility tensor dependence on the lattice deformations, i.e.~those caused by excited vibrational modes. Considering tiny displacements out of the equilibrium position, the susceptibility can be perturbatively expanded around its initial value $\boldsymbol{\chi}^{(0)}$ as a function of the lattice normal modes coordinates $q_n$, also referred to as phonon positions ($n$ labels the mode) \cite{bib:cars, bib:Boyd}:
\begin{equation}
\begin{split}
\chi_{\lambda\lambda’}(q_1, \dots , q_N) &= \chi^{(0)}_{\lambda\lambda’} + \sum_{n}\chi^{(1)}_{\lambda\lambda’}(n) q_{n}.\\
\end{split}
\label{eq:pol(q)}
\end{equation}
where we defined $\chi^{(1)}_{\lambda\lambda’}(n) := {{\Big({\delta\boldsymbol{\chi}}/ {\delta q_n}\Big)}_{\lambda\lambda’} \big|_{q_n= 0}}$ the components of the rank three non-linear susceptibility tensor $\boldsymbol{\chi}^{(1)}$ and $\lambda$ is the polarization index.\\
In order to simplify the notation, in the following we neglect the summation over $n$ and discuss the interaction of a single phononic mode with light. The characteristic structure of different modes will be highlighted in the last part of the paper, where we consider the specific case of quartz and compare the experimental evidences with the model predictions. 

The refractive index depends on the susceptibility, $ \boldsymbol{n} = \sqrt{1+\boldsymbol{\chi}}$. This result in a modulation of the material refractive properties as a function of the phonon position operator to be introduced below\footnote{In this paper we only take into account the refractive effects involving the transmitted light fields. We note that extending the presented formalism to the reflective degrees of freedom also their response can be treated.}.
  
The explicit form of the bulk hamiltonians is obtained from the energy density required to establish the polarization $\vec{P}$ in a dielectric sample, which is given by~\cite{bib:Boyd}
\begin{equation} \label{energy}
\begin{split}
U(\vec{x},t) =&  -  \vec{P} (\vec{x},t) \cdot \vec{E}(\vec{x},t)\\=& - \epsilon_0 \sum_{\lambda\lambda'} \chi_{\lambda\lambda’} {E_{\lambda'}}(\vec{x},t) E_{\lambda}(\vec{x},t).
\end{split}
\end{equation}\\
From this expression, substituting the susceptibility as in \eqref{eq:pol(q)} and quantizing the electric field, we can single out two main contributions to the photo-phonon interaction (see Supplementary Material) \cite{bib:Andrea}.\\
The first term, which we dub \textit {refractive }, is given by:
\begin{equation}
\begin{split}
\label{polproc}
H_{Ref}=-\frac{V_S}{2V}\sum_{\lambda\lambda',j}\omega_j(\chi^{(0)}_{\lambda\lambda'}+q\chi^{(1)}_{\lambda\lambda'})\Big(a^\dag_{\lambda j}\,a_{\lambda' j}\,+\,a_{\lambda j}\,a^\dag_{\lambda' j}\Big)\\
\end{split}
\end{equation}
where  $V_S$ and $V$ are the sample and quantization volumes and $\omega_j$ the photon frequencies indexed by $j$. $H_{Ref}$ describes the redistribution of photons between the two polarizations which is mediated by the static birefringence ($\boldsymbol{\chi}^{(0)}$) and the time dependent contribution  ($\boldsymbol{\chi}^{(1)}$). The latter being ruled by the instantaneous atomic position as highlighted in \eqref{eq:pol(q)}.

The second term contributing to the hamiltonian, dubbed \textit {Raman}, is given by:
\begin{equation}
\label{Ram1}
\begin{split}
H_{Ram}=- \frac{\sqrt{V_S}}{2V\sqrt{2m \Omega}}\sum_{\lambda \lambda', j}\omega_j  \, {\chi}^{(1)}_{\lambda \lambda'}\Big[\Big(\,a^\dag_{\lambda j}a_{\lambda' j+\frac{\Omega}{\delta}}\Big)\,b^\dag\,\\
+ \,\Big(\,a_{\lambda j}a^\dag_{\lambda' j+\frac{\Omega}{\delta}}\Big)\,b\Big],
\end{split}
\end{equation}

where $\Omega$ is the phonon frequency and $m$ its effective mass. 
Considering the photonic ($a$) and phononic ($b$) ladder operators, in the two terms of $H_{Ram}$ represent the Stokes and Anti-Stokes. Photons with energy $\omega_j$ and polarization $\lambda$ are destroyed by $a_{\lambda j}$ and photons of energy $\omega_j\pm\Omega$ and polarization $\lambda'$ are created by $a^\dag_{\lambda' j\pm\frac{\Omega}{\delta}}$, together with the emission ($b^{\dagger}$) and annihilation ($b$) of a phonon, respectively.

We stress that $H_{Ref}$ and $H_{Ram}$ are representative of the major effects observed in experiments. In particular, $H_{Ref}$ acts as a beamsplitter relocating photons at a fixed frequency between the two polarizations, that does not imply an effective transfer of energy between the light and the sample. Conversely, $H_{Ram}$ involves the exchange of a quantum of the elastic energy between the light pulse and the crystal, which results in a transfer of spectral weight between different spectral components.

The system dynamics is obtained acting with the combination of $H_{Ref}$ and $H_{Ram}$, considering the pulse-sample interaction time ($\tau$) shorter than the phonon oscillation period.

Concerning the initial states, we describe the impinging light pulse as a multimode coherent state $\vert\alpha\rangle$, where $\alpha$ stands for the vector with components $\alpha_{\lambda j}$, given by
\begin{equation}
\label{Probe1}
\vert\alpha\rangle=\exp\Big(\sum_{\lambda, j}\alpha_{\lambda j} a_{\lambda j}^\dag-\alpha_{\lambda j}^* a_{ \lambda j}\Big)\vert 0\rangle\ ,\quad
a_{\lambda j}\vert\alpha\rangle=\alpha_{\lambda j}\,\vert\alpha\rangle\ ,
\end{equation}
with annihilation and creation operators of photonic modes $a_{\lambda j}$ and $a^\dag_{\lambda j}$ such that
$ \Big[a_{\lambda j}\,,\,a^\dag_{\lambda'k}\Big]=\delta_{jk}\delta_{\lambda\lambda'}$,
where $\vert 0\rangle$ is the vacuum state and $\lambda ,j$ are the polarization and frequency indices, respectively. In particular, we consider a set of modes centered around the frequency $\omega_0$ and spaced by $\delta$: $\omega_j = j\delta+ \omega_0$. 

The phononic degree of freedom, (we provide the general model considering only one vibrational mode), is described through the creation and annihilation operators $b$ and $b^\dag$, satisfying the commutation relation $[b,b^\dag]=1$. Accordingly, the position and momentum phonon operators are defined as linear combinations of $b$ and $b^\dag$:
\begin{equation}\label{defpq}
q= \frac{1}{\sqrt{2m\Omega V_S}}(b+b^\dag), \qquad p= \sqrt{\frac{m\Omega}{2V_S}}i(b^\dag - b),
\end{equation}
where $\Omega$ is the frequency of the mode, $m$ is the effective mass and $V_S$ is the volume of the sample. 
Using this notation, in the following we discuss separately the two different effects, LRM and ISRS, commenting on how they modify the transmitted light and the phononic phase-space.

\subsection{Refractive modulation}

$H_{Ref}$, \eqref{polproc}, describes the redistribution of photons between the two polarizations.
\\
We exploit the first order expansion of the matrix $\boldsymbol{\chi}$ in \eqref{eq:pol(q)} to split the refractive hamiltonian in an equilibrium term
\begin{equation}
\begin{split}
H_{Ref}^{(0)}&=-\frac{V_S}{2V}\sum_{\lambda\lambda',j}\omega_j\chi^{(0)}_{\lambda\lambda'}\Big(a^\dag_{\lambda j}\,a_{\lambda' j}\,+\,a_{\lambda j}\,a^\dag_{\lambda' j}\Big),\\
\end{split}
\end{equation}
and a dynamical one
\begin{equation}
\begin{split}
H_{Ref}^{(1)}&=-\frac{V_S}{2V}\,q\sum_{\lambda\lambda',j}\omega_j\chi^{(1)}_{\lambda\lambda'}\Big(a^\dag_{\lambda j}\,a_{\lambda' j}\,+\,a_{\lambda j}\,a^\dag_{\lambda' j}\Big).\\
\end{split}
\end{equation}
 $\boldsymbol{\chi}^{(0)}$ is the equilibrium susceptibility, that is without phonon excitation, which describes static refractive effects like polarization rotation and birefringence. In particular, we consider the case of an isotropic sample with an hermitean suscepitibility of the form
 \begin{equation}
 \label{chi0}
 \boldsymbol{\chi}^{(0)} = \begin{pmatrix}
u & |w|e^{i\phi}\\
|w|e^{-i\phi}&u
\end{pmatrix}
 \end{equation}
 where $|w|$ and $\phi$ quantify respectively the polarization rotation and ellipticity induced in a linearly polarized input beam.\\
 The phonon related non-linear susceptibility coefficients $\chi^{(1)}_{\lambda\lambda'}$ are assumed real, such that $\chi^{(1)}_{\lambda\lambda'}=\chi^{(1)}_{\lambda'\lambda}$, and small in absolute value, so they represent a perturbative modification of the equilibrium tensor. 

In the experimental realization the polarization time domain changes are measured with an analyzer aligned along a reference frame defined in order to compensate the equilibrium rotation from the sample. In the model, we account for this with an additional term in the hamiltonian (Supplementary Material).\\



\subsection{Impulsive Stimulated Raman Scattering}

The energy modulation in the sample is modelled by the Raman Hamiltonian, $H_{Ram}\,$ \eqref{Ram1}.

\begin{figure} [htbp]
\centering
\mbox{\includegraphics[width=8cm]{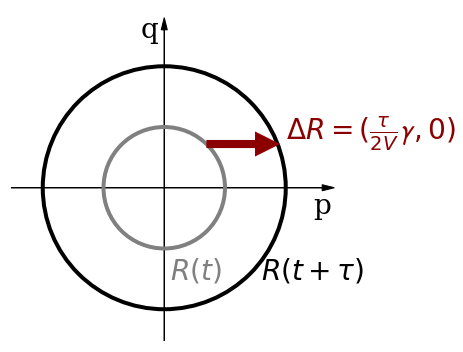}} 
\caption{Phase space representation of the Raman interaction between light and phonon. The circular trajectory is the one followed by the free evolution of the coherent phonon, modelled as an harmonic oscillator. Light-phonon imparts a positive momentum displacement (arrow) modifying the radius of the phonon trajectory.}
\label{fig:phasespace}
\end{figure}

The evolution of the phononic operator $b$ reads
\begin{eqnarray}\label{b}
b(\tau) =  b(0) +i\frac{\tau\sqrt{V_S}}{2V\sqrt{2m \Omega}}  g , \\ \textrm{where} \quad g= \sum_{\lambda \lambda',j} {\chi}^{(1)}_{\lambda \lambda'}\, \omega_j\, a^\dag_{\lambda j} a_{\lambda' j+\frac{\Omega}{\delta}}.\nonumber
\end{eqnarray}
which in turn gives mean values of the phonon phase-space variables position $q$ and momentum $p$, modified with respect to a generic initial state as 
\begin{eqnarray}
\label{phsp}
&\begin{cases}
\braket{q(\tau)}= \braket{q(0)},\\
\braket{p(\tau)}= \braket{p(0)}+\frac{\tau}{2V}\gamma,
\end{cases}\\
&\textrm{with}\quad \gamma=\braket{g} = \sum_{\lambda \lambda',j}\chi^{(1)}_{\lambda \lambda'}\, \omega_j \, |\alpha_{\lambda j}| |\alpha_{\lambda' j+\frac{\Omega}{\delta}}| .\nonumber
\end{eqnarray}
This shows that the result of a sudden Raman interaction is a displacement along the momentum axis, as depicted in figure \ref{fig:phasespace}. The squared radius $R^2$ gives the mean value of the phonon number $N=b^\dag b$, which, to second order in the $\tau \boldsymbol{\chi}^{(1)}$ coupling parameter, results
 \begin{equation}
\braket{N(\tau)}= \braket{N(0)} +  \frac{\tau V_S}{2Vm\Omega}\gamma \braket{p(0)} + \frac{\tau^2V_S}{8V^2m\Omega} \braket{g^\dag g}.
\end{equation}

We notice that the first order contribution depends on the value of the momentum $p$ before the interaction, while the second order term is proportional to the mean value of the operator $g^\dag g$, which equals $\gamma^2$ if light states are classical (coherent states such that $|\alpha|^2 \gg 1$). The second order term is usually negligible with respect to the first one unless $\braket{p(0)}=0$. 

The effects on the phononic degrees of freedom have their counterparts on the photonic ones. 
The intensity of the transmitted light at a certain frequency $\omega_j$ and polarization $\lambda$, computed as $\braket{I_{\lambda j}(\tau)}:=\braket{a^{\dag}_{\lambda j}(\tau)\,a_{\lambda j}(\tau)}$ reads 
\begin{equation}
\label{isrsint}
\begin{split}
\braket{I_{\lambda j}(\tau)}=&\braket{I_{\lambda j}(0)}\\&+ \frac{\tau V_S}{2Vm\Omega} \sum_{\lambda’} {\chi}^{(1)}_{\lambda\lambda’}\omega_j |\alpha_{\lambda j}|\\&\times \Big(|\alpha_{\lambda’ j+\frac{\Omega}{\delta}}|-|\alpha_{\lambda’ j-\frac{\Omega}{\delta}}| \Big)\left(\braket{p(0)}+\frac{\tau}{4V} \gamma \right) \\&+\tau^2 {\gamma'}_j.
\end{split}
\end{equation}
In \eqref{isrsint} the term in $\braket{p(0)}$ results from first order contributions and is proportional to the difference in amplitude between the modes corresponding to the frequencies $\omega_j + \Omega$ and $\omega_j - \Omega$. The terms in $\gamma$ and $\gamma'$ result from second order interaction. Among them one can recognize a contribution with a structure similar to the first order ($\braket{p(0)}$ substituted by $\gamma$) and a further one $\gamma'_j$ which depends on the mean-values of squared phonon operators (see  Supplementary Material for $\gamma'$ full derivation).
 
Equipped with this general machinery, we now proceed to study in detail the quantum signatures in pump\&probe experiments.

\section{Pump and probe approach}
\label{sec: pumpprobe}

\begin{figure*} [htbp]
\centering
\includegraphics[width=16cm]{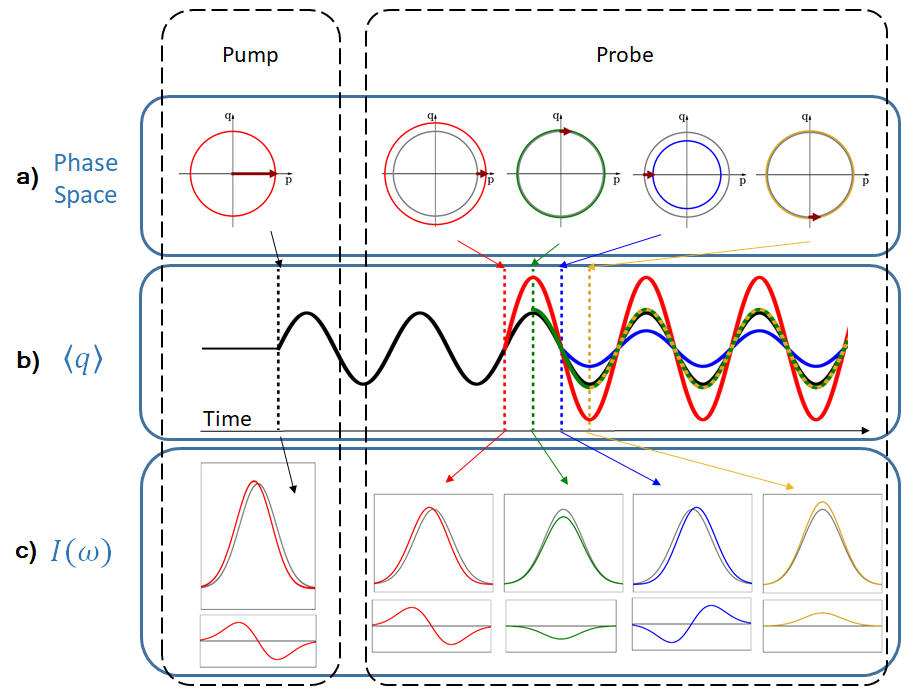}
\caption{Summary of the predicted effects. a) Pump and probe induced displacements (arrow) describe the effect on the vibrational energy ($\frac{\braket{p^2}}{2m}+\frac{1}{2}m\Omega^2\braket{q^2}$). Depending on the phase space coordinates at the interaction time the phonon oscillation (b) can be amplified (red) or damped (blue). The corresponding effect on the transmitted pulse spectra (c) is a \emph{red-shift} or \emph{blue-shift}, respectively. The energy exchange is most important at the momentum extremes. At the extremal positions the oscillation is negligibly amplified and the relevant effect is a modulation of the transmittivity which does not change the spectral content (green and gold).      \label{PhaseSpace}}
\end{figure*}

Pump\&probe experiments provide standard techniques in time-resolved spectroscopy, whereby a first intense laser pulse (the pump) excites the vibrational degrees of freedom of a sample and a second pulse, less intense, is used to probe non-equilibrium features. By repeating the experiment at different time-delays between pump and probe, one can retrieve information about the phonon dynamics in the sample.\\
In the following, we describe how the theoretical model presented in the previous section applies in this framework, highlighting the different effects due to the pump and the probe pulses. We will consider the pump acting on the phononic equilibrium state at a reference time $t = 0$, and study the probe response as a function of the delay time $t$. In particular, we focus on frequency and polarization resolved intensity measurements, that we can describe through the LRM and ISRS effects. Figure \ref{PhaseSpace} shows a sketch of pump\&probe interactions with reference to the phonon phase space description.

\subsection{Pump-target interaction}
We assume the pump impinging on the sample at equilibrium, where the phononic position and momentum have zero average $\braket{q(0)}=\braket{p(0)}=0$ (figure \ref{PhaseSpace}, left). This is the case for instance if the initial state of the vibrational degrees of freedom has a thermal distribution. \\
The ISRS effect on the intensities of the different frequency components of the pump pulses are here evaluated neglecting the equilibrium LRM. The first order term is null because of $\braket{p(0)}=0$ and we also neglect the term ${\gamma'}_j^{pump}$ because the phonon population is negligible with respect to the photon number. The transmitted pump intensity is given by
\begin{equation}\label{intpump}
\begin{split}
&\braket{I^{pump}_{\lambda j}(\tau)}_0 
=\braket{I^{pump}_{\lambda j}(0)}_0\\&+\displaystyle\scriptstyle\frac{\tau^2V_S}{8V^2m\Omega} \displaystyle{\gamma}^{pump} \sum_{\lambda’} \chi^{(1)}_{\lambda\lambda’}\omega_j |{\alpha}^{pump}_{\lambda j}|\Big(|{\alpha}^{pump}_{\lambda’ j+\frac{\Omega}{\delta}}|-|{\alpha}^{pump}_{\lambda’ j-\frac{\Omega}{\delta}}| \Big),
\end{split}
\end{equation}
where $\braket{I^{pump}_{\lambda j}(0)}_0$ is the intensity of the spectral components priorly to the interaction and the second term can be interpreted as an effective red-shift of the pulse spectrum. Indeed, assuming the incoming pulse to have a Gaussian spectrum centered in $\omega_0$, equation \eqref{intpump} implies that modes with frequency smaller that $\omega_0$ are amplified (because the difference $|{\alpha}^{pump}_{\lambda’ j+\frac{\Omega}{\delta}}|-|{\alpha}^{pump}_{\lambda’ j-\frac{\Omega}{\delta}}| $ is positive), while modes with frequency higher than $\omega_0$ are suppressed. This description rationalizes well the pump red-shift observed in experiments \cite{bib:ISRSNelson}. \\
Correspondingly, according to \eqref{phsp}, the phonon system is shifted from the origin of the phase space ($\braket{q(0)}_0= 0$,$\braket{p(0)}_0= 0$) along the momentum axis to a trajectory of radius 
\begin{equation}\label{phon}
R \equiv \braket{p(\tau)}_0 = \frac{\tau}{2V} {\gamma}^{pump} .
\end{equation}

\subsection{Probe-target interaction}
We consider the probe interacting at an arbitrary delay-time $t$. Considering that the excitation starts when the pump interacts ($t = 0$), at negative times the probe still sees the system in equilibrium. Afterwards, the considered phonon mode is excited and the atomic positions oscillate at the corresponding phonon frequency $\Omega$. In the following, we neglect dissipative effects occurring in the time-interval between the action of the two pulses, the evolution being described by the Hamiltonian of a free quantum harmonic oscillator. As a consequence, the initial conditions for the probe interaction at a given time delay ($t$) are: 
\begin{equation}
\begin{cases}
t<0, & \braket{q(0)}_t= \braket{p(0)}_t=0\\
t>0, & \braket{q(0)}_t= \frac{R}{m\Omega} \sin(\Omega t) ,\quad \braket{p(0)}_t = R\cos(\Omega t).
\end{cases}  
\end{equation}
In the following we consider only positive delay-times $t>0$ (figure \ref{PhaseSpace}, right dashed box) and estimate the dynamical intensity modulation $\braket{\Delta I^{probe}_{\lambda j}}_t = \braket{I^{probe}(\tau)_{\lambda j}}_t - \braket{I^{probe}(\tau)_{\lambda j}}_{<0}$ with respect to the unperturbed condition (negative times).\\
Applying $H_{Ref}$ up to first order in $\tau\boldsymbol{\chi}^{(1)}$, the dynamical response due to the LRM effect is dependent on the phonon position and reads
\begin{equation}
\label{probeintref}
\begin{split}
&\braket{{\Delta I^{probe}_{\lambda j}}_{Ref}}_t = \\ 
&-\frac{V_S}{2V} \sum_{\lambda'\lambda''}\big(\boldsymbol{K}(\tau,\boldsymbol{\chi}^{(0)},\boldsymbol{\chi}^{(1)})\big)_{\lambda, \lambda' \lambda''}|\alpha_{\lambda'j}||\alpha_{\lambda''j}| \braket{q(0)}_t,
\end{split}
\end{equation}
where all the terms which take account of the equilibrium refractive properties are collected in the $\boldsymbol{K}$ tensor (these are explicitely calculated in the Supplementary Materials). Eq.\eqref{probeintref} represents an intensity modulation of the trasmitted light which is modified by the instantaneous position of the atoms. \\ 
For the ISRS, \eqref{isrsint}, contribution only, we obtain an explicit expression for the resulting intensity as a function of phonon momentum,
\begin{equation}
\label{probeintram}
\begin{split}
&\braket{{\Delta I^{probe}_{\lambda j}}_{Ram}}_t 
= \\ 
&+ \scriptstyle\frac{\tau V_S}{2Vm\Omega} \displaystyle\sum_{\lambda’} \chi^{(1)}_{\lambda\lambda’}\omega_j |{\alpha}^{probe}_{\lambda j}|\Big(|\alpha^{probe}_{\lambda’ j+\frac{\Omega}{\delta}}|-|{\alpha}^{probe}_{\lambda’ j-\frac{\Omega}{\delta}}| \Big)\braket{p(0)}_t , 
\end{split}
\end{equation}
where we have neglected the second order ISRS terms considering that $\gamma^{probe} \ll \braket{p(0)}_t  $ due to $|\alpha^{pump}| \gg |\alpha^{probe}|$. In the following section and in the Supplementary Materials also the combined action of Raman ($H_{Ram}$) and refractive equilibrium effects ($H_{Ref}^{(0)}$) is considered.\\ 
We will use these expressions as a benchmark for the model, comparing the predicted results with experimental data for the probe transmitted intensity. In figure \ref{PhaseSpace} the expected probe spectral modulations are highlighted for the phonon position and momentum extremes.


\section{Case study: quartz}
\label{sec: quartz}

In the model developed so far, we considered only one phononic mode of frequency $\Omega$. The generalization to many phononic modes is straightforward. However, since we are interested only in the first order corrections to the transmittivity of the probe pulse, we can add the contribution of different phononic modes independently. In other words, the coupling between different phononic modes would be a higher order effect.\\
We now discuss the case study of $\alpha$-quartz excited along its $c$-axis \cite{bib:Porto, bib:Wefers}. In this setting, three different symmetry classes for the phononic modes come into play. Their different contributions can be selected by a proper combination of the pump-probe polarization \cite{bib:Rundquist}. These three classes correspond to specific properties of the susceptibility tensor. In particular, for the classes called $A$ (totalsymmetric), and the two degenerate $E_L$ (longitudinal) and $E_T$ (transverse), $(\chi^{(1)}_n)_{\lambda\lambda’}$ have the form 
\begin{equation}
A =
\begin{pmatrix}
a&0\\
0&a
\end{pmatrix}, \ E^L = 
\begin{pmatrix}
c_L&0\\
0&-c_L
\end{pmatrix}, \ E^T =
\begin{pmatrix}
0&-c_T\\
-c_T&0
\end{pmatrix}.
\end{equation}
In the following, by varying the angle between the pump and the probe polarization, we discuss the transmittivity dependence on the symmetry classes.

\subsection{Model prediction}\label{modpred}

In order to make quantitative predictions, we fix specific features of the pulses.
Both pump and probe are chosen to be linearly polarized laser beams with a Gaussian spectrum of height $\alpha_0>0$
\begin{equation}
\label{amplitudes}
\alpha_{ j}=\alpha_0\,{\rm e}^{-(j\delta)^2/(2\sigma^2)}\, ,
\end{equation}
where $\sigma$ is the width of the pulse frequency distribution. The difference in intensity between pump and probe is accounted for by setting $|\alpha_0^{pump}| \gg |\alpha_0^{probe}|$.
 We consider a reference frame such that the probe is initially polarized along the $x$ axis while the pump is oriented at an angle $\theta$ with respect to it.
In order to make the dependence on $\theta$ explicit, we choose the initial state of the pump such that $a_{\lambda j} | \alpha^{pump} \rangle= \alpha^{pump}_{\lambda j} | \alpha^{pump} \rangle$ where
\begin{equation}
\alpha^{pump}_{x j}= \alpha^{pump}_{ j} \cos(\theta)  , \quad \alpha^{pump}_{y j}= \alpha^{pump}_{ j} \sin(\theta).
\end{equation}

The transmittivity of the probe after the action of the pump depends on the radial parameter $R$ introduced in \eqref{phon}, which, according to our model, contains all the information about the phonon dynamics. In particular, it turns out that
\begin{equation}
\label{polartrend}
\begin{split}
R_A = &\,a {\eta}^{pump}_{\Omega_A} ,\\ R_{E^L} =&\, c_{L}\ \cos(2\theta) \ {\eta}^{pump}_{\Omega_E} , \\ R_{E^T} =& -c_{T}\ \sin(2\theta) \ {\eta}^{pump}_{\Omega_E} , 
\end{split}
\end{equation}
where the parameter ${\eta}_{\Omega}^{pump}$ has been defined as follows 
\begin{equation}
{\eta}^{pump}_{\Omega}=  \frac{\tau}{2V}\sum_j \omega_j |{\alpha}^{pump}_{j}| |{\alpha}^{pump}_{j+\frac{\Omega}{\delta}}|.
 \end{equation}
 The final expressions for the dynamical modulation of transmitted intensity are simplified considering a small equilibrium rotation, i.e. considering up to linear order in the $\boldsymbol{\chi}^{(0)}$ coefficient $|w|$, \eqref{chi0}. 
When the analyzer selects the polarization along the $x$ axis, the leading contributions are the ISRS one (zero-th order in $|w|$)
\begin{equation}
\begin{split}
\label{x}
&\braket{\Delta I^{probe}_{x j}(\tau)}_t = 
+ \scriptstyle\frac{\tau V_S}{2V}\displaystyle\omega_j |{\alpha}^{probe}_{j}|\\ &\times\Big[a^2 \left(|{\alpha}^{probe}_{j+\frac{\Omega_A}{\delta}}|-|{\alpha}^{probe}_{j-\frac{\Omega_A}{\delta}}|\right) \frac{{\eta}^{pump}_{\Omega_A}}{m_A\Omega_A}\cos(\Omega_A t)\\&
\,+c_L^2 \cos(2\theta) \left(|{\alpha}^{probe}_{j+\frac{\Omega_E}{\delta}}|-|{\alpha}^{probe}_{j-\frac{\Omega_E}{\delta}}|\right) \frac{{\eta}^{pump}_{\Omega_E}}{m_E\Omega_E}\cos(\Omega_E t)\Big] ,
\end{split}
\end{equation}     
while, choosing the polarization along the $y$ axis the relevant terms are linear in $|w|$
\begin{equation}
\label{y}
\begin{split}
&\braket{\Delta I^{probe}_{y j}(\tau)}_t =\\&+\scriptstyle\frac{\tau V_S}{2V}\displaystyle\frac{c_T^2 {\eta}^{pump}_{\Omega_E}}{m_E \Omega_E} \tau |w| \sin(2\theta)(1-\cos\phi)|{\alpha}_j^{probe}|\\
&\times \big[2|{\alpha}_j^{probe}|+ \Big(|{\alpha}_{j+\frac{\Omega_E}{\delta}}^{probe}|+|{\alpha}_{j-\frac{\Omega_E}{\delta}}^{probe}|\Big)\big]\sin(\Omega_E t).\\
\end{split}
\end{equation}
The first term in \eqref{y} is relative to the phonon dependent refraction (i.e. ruled by $H_{Ref}^{(1)}$); the second is due to the combined action between ISRS ($H_{Ram}$) and LRM equilibrium refraction ($H_{Ref}^{(0)}$). The resulting intensity modulation evolves in phase with the phonon position and has the same sign for every frequency.
 
We summarize here some basic conclusions that can be drawn from the previous equations and which are compatible with the existing classical descriptions present in the literature \cite{bib:Merlin0, bib:RevNelson}:
\begin{itemize}
\item Different modes of vibration in the crystal can be selectively measured by appropiately choosing the polarization of the pump and probe pulses. 
\item The LRM effect does not couple different modes of light and produces a global amplitude modulation \eqref{probeintref}, whereas the Raman process gives a shift in the spectral weight, preserving the total intensity \eqref{isrsint}.
\item For a given phonon, the modification of the transmittivity produced by the displacement dependent LRM effect oscillates in time with a different phase with respect to the momentum dependent ISRS shift effect. In particular, when the Raman effect is maximum the refractive modulation is zero and viceversa, so that it is possible to distinguish the two processes looking at specific time-delays between pump and probe.
\end{itemize}

\subsection{Experiment}

In pump\&probe experiments we measure the modulation of the probe transmitted intensity $\braket{\Delta I^{probe}_{\lambda j}}_t = \braket{I^{probe}(\tau)_{\lambda j}}_t - \braket{I^{probe}(\tau)_{\lambda j}}_{<0}$, as a function of time delay and probe frequency. The experimental outcome can thus be compared with the theoretical prediction obtained by means of the expressions \eqref{x} and \eqref{y}. Furthermore, by adjusting the experimental parameters, one can select the pump orientation $\theta$ and the analyzed polarization $\lambda$. The experimental setup details are given in the supplementary material \cite{bib:Filippo}. The employed pulse fluences are 0.8 mJ/cm$^{2}$ for the pump and 0.7 $\mu$J/cm$^{2}$ for the probe.

In this section we present two peculiar configurations, the particular geometry of which is useful to discuss and verify the main predicted features and distinguish LRM and ISRS effects. We present the data normalized by the unperturbed (negative times) peak intensity $\braket{I^{probe}(\tau)_{x 0}}_{<0}$. We obtain the phonon frequencies performing the Fourier Transform (FT) of the positive delays and we focus on the spectral shape of the modulation at relevant times. We set the zero delay at the centre of the overlap between pump and probe pulses (pulse duration $\simeq$40 fs)\footnote{Notice that around this region the interpretation is complicated by interference and other effects related to the pulse duration.}.\\
In figure \ref{Spectral_par} we report the measurement obtained with the pump polarized along the $x$ axis, $\theta = 0^\circ$, and the transmitted light is measured in the parallel polarization. We notice in the phonon spectrum the presence of both $A$ and $E$ symmetry peaks and we highlight the ISRS red/blue-shift modulations. This is consistent with the thoeretical model \eqref{x}. 
\begin{figure} [htbp]
\centering
\includegraphics[width=8cm]{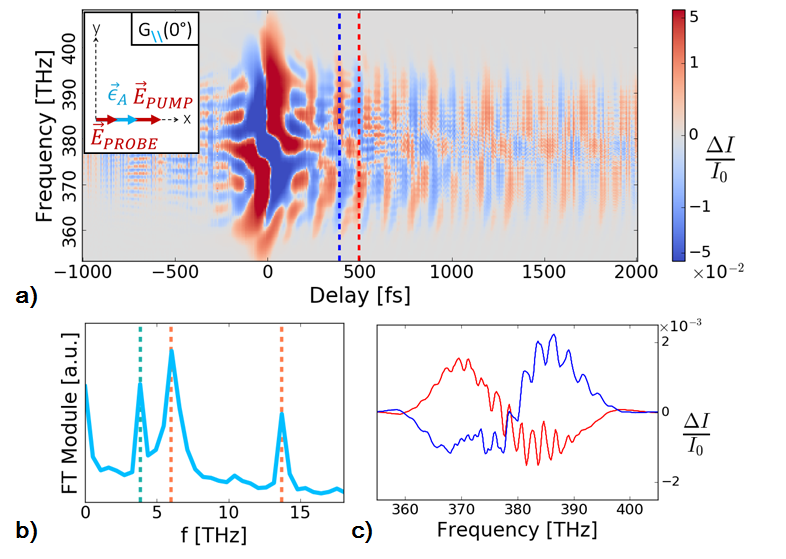}\\
\caption{Results depending on the relative orientation between pump and probe polarizations and analyzer direction. a) Spectral modulation vs delay in the $(\lambda = x, \, \theta=0^\circ)$ configuration (insert) is presented. b) FT: 4 THz $E_L$ phonon is detected, together with 6 THz and 14 THz $A$ symetry modes. c) The transmittivity modulation is selected at $t = 391$~fs (blue) and $t = 498$~fs (red). Spectral shifts resulting from ISRS are observed. \label{Spectral_par}}
\end{figure}\\
In figure \ref{Spectral_perp}, we keep the probe oriented along $x$, while the pump is rotated $\theta = 45^\circ$ and the analyzer is set cross ($y$ axis). We observe modulations of the same sign along the full spectrum. The model \eqref{y} correctly predicts that only the oscillation produced by the transverse mode is visible in the FT.\\ A comparison between figure \ref{Spectral_par} and \ref{Spectral_perp} confirms the phase shift between ISRS and LRM predicted by the model.
\begin{figure} [htbp]
\centering
\includegraphics[width=8cm]{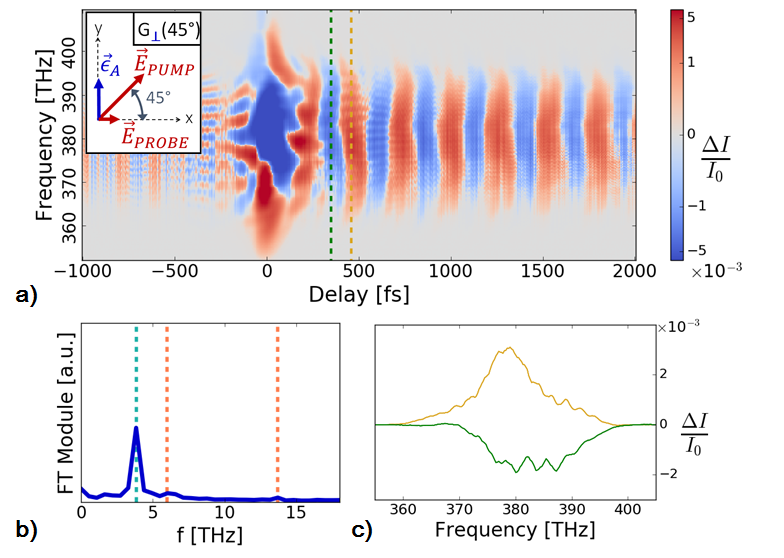}\\
\caption{a) Spectral modulation vs delay in the $(\lambda = y, \, \theta=45^\circ)$ configuration (insert) is presented. b) FT: only the 4 THz $E_T$ phonon is detected. c) The transmittivity modulation is selected at $t = 351$~fs (green) and $t = 458$~fs (gold). As expected for the refractive modulation the transmittivity change has the same sign for the whole spectrum. \label{Spectral_perp}}
\end{figure}

\begin{figure} [htbp]
\centering
\mbox{\includegraphics[width=8cm]{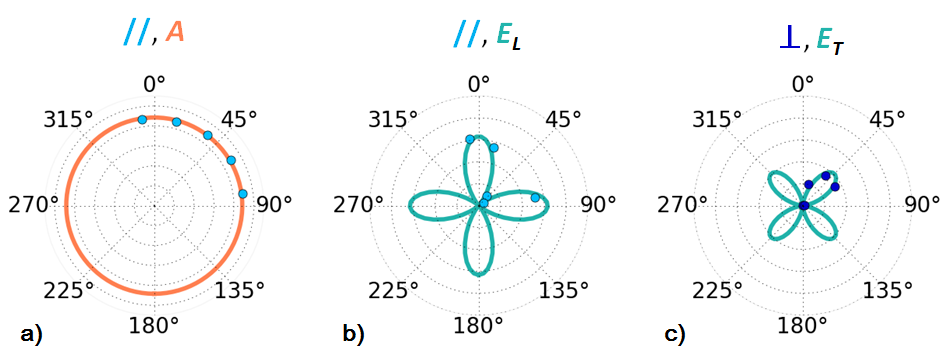}} 
\caption{Polar plots representative of the different phonon symmetries are presented for an analyzer setted parallel (a,b) or orthogonal (cross) (c) to the probe polarization. Measured signal intensity (dots) for each considered mode is fitted (line) with the proper dependence on relative orientation between pump\&probe, suggested by \eqref{polartrend}. a) $A$-mode 6 THz, b) $E^L$-mode 4 THz phonon, parallel polarization. c) $E^T$-mode 4 THz phonon, cross polarization.\label{PolarPlots}}
\end{figure}

In addition to these configurations, we verify the symmetry properties by means of measurements for other values of $\theta$.
In figure \ref{PolarPlots} we show summary polar plots where we represent the peak intensity of the FT for a selected phonon as a function of the pump orientation and the selected polarization.  


\section{Conclusions}

In conclusion, we presented a theoretical quantum model that describes the interactions between the vibrational degrees of freedom of a crystal (phonons) and a multimode light beam. Two different effects, not clearly distinguished in the literature, have been studied, namely the Impulsive Stimulated Raman Scattering and the Linear Refractive Modulation. The model is applied to the description of pump\&probe  experiments, where a first intense laser pulse is used to excite the vibrations in a sample and then a second light beam is used to probe the dynamics of these vibrations, highlighting the different features in the two situations. In particular, for experiments based on $\alpha$-quartz, in different geometrical settings the theoretical predictions agree with the experimental findings. Though the results are compatible with classical descriptions of light-matter interaction, the quantum model can be applied to  more general situations where classical models fail. In this respect, this paper should be seen as a first benchmark on the quality of our model which can be further exploited in the context of different measurements unveiling purely quantum effects. Moreover, it can be used as a tool to infer the non-equilibrium properties of complex materials via spectroscopic measurement.

\ack
This work was supported by the European Research Council through the project INCEPT (grant Agreement No. 677488).\\
F.G., G.S., M.E. and D.F. worked on the pump-probe experiment with quartz, while F.G., S.M., A.B. and F.B. developed the theoretical model.\\
F.G. and S.M. contributed equally to this work.


{\scriptsize
\bibliographystyle{unsrt}
}

\section*{References}


\onecolumn

{\centering \bf \Large Supplementary material: \\ Quantum model for Impulsive Stimulated Raman Scattering}

\section{Quantum model}
A full quantum description of the pulse-target photon-phonon interaction can be provided by assuming the process to consist of two independent, that is dynamically decoupled contributions:
\begin{enumerate}
\item
a phonon-dependent refraction process that rotates the polarization of the incoming laser beam (we called this process Linear Refractive Modulation, LRM, in the main text);
\item
Raman Stokes and anti-Stokes processes affecting the transmitted photons while they cross the target and interact with its vibrational degrees of freedom (Impulsive Stimulated Raman Scattering, ISRS, in the main text).
\end{enumerate}
These processes act on an incoming mode-locked laser pulse consisting of linearly polarized photons with frequencies $\omega_j=\omega_0+j\delta$, $-J\leq j\leq J$, described by a coherent state
\begin{equation}
\label{coherent}
\vert\alpha\rangle=\exp\Big(\sum_{j, \lambda}\alpha_{\lambda j} a_{\lambda j}^\dag-\alpha_{\lambda j}^* a_{\lambda j}\Big)\vert 0\rangle\ ,\quad
a_{\lambda j}\vert\alpha\rangle=\alpha_{\lambda j}\,\vert\alpha\rangle\ ,
\end{equation}
with annihilation and creation operators $a_{\lambda j}$ and $a^\dag_{\lambda j}$ such that
$ \Big[a_{\lambda j}\,,\,a^\dag_{\lambda'k}\Big]=\delta_{jk}\delta_{\lambda\lambda'}$,
where $\vert 0\rangle$ is the vacuum state, $\lambda=x,y$ are polarization indices and $j$ a frequency index.
The $N=2J+1$ frequencies $\omega_j = \omega_0 + j \delta$ are distributed  around a central frequency $\omega_0$ with $\delta$ a constant depending on the laser repetition rate.
Furthermore, we will consider an $\omega_0$ centered, Gaussian shaped pulse so that, together with the mode-locking condition on the phases of the contributing amplitudes ($\varphi_j = j\varphi'+\varphi_0$),
\begin{equation}
\label{amplitudes}
\alpha_{\lambda j}=\alpha_\lambda\,{\rm e}^{-(j \delta)^2/(2\sigma^2)}\,{\rm e}^{i\,(j\, \varphi'+\varphi_0)}\ ,
\end{equation}
where $\alpha_{\lambda}>0$ and $\varphi'$ is the mode-locking reference phase (that is assumed to be zero in the following) and $\sigma$ is the width of the frequency distribution in the pulse.

For sake of simplicity, the vibrational properties of the target will be described by a single phonon mode of energy $\Omega$ associated with bosonic annihilation and creation operators $b$ and $b^\dag$ such that $[b\,,\, b^\dag]=1$.

\subsection{Physical justification of the model}

In the following we present some considerations in order to physically justify the model we are considering. In particular, the Raman interaction Hamiltonian and the bulk rotation are two contributions related to the dipole energy density that reads 
 \begin{equation}
 U^{\mathrm{int}}(\vec{x})= -\epsilon_0 \sum_{\lambda \lambda'} \chi_{\lambda \lambda'} E_{\lambda}(\vec{x}) E_{\lambda'}(\vec{x}) = -\epsilon_0 \sum_{\lambda \lambda'}\left( \chi^{(0)}_{\lambda \lambda'} + \sum_n \chi^{(1)}_{n; \lambda \lambda'} q_n (\vec{x}) \right) E_{\lambda}(\vec{x}) E_{\lambda'}(\vec{x}) ,
 \end{equation}
 where the elastic field $q_n (\vec{x})$  describing the vibration in the crystal, together with its momentum $p_n (\vec{x})$, and the electric field $E_{\lambda}(\vec{x})$ polarized along $\lambda$ can be quantized as follows
 \begin{equation}
 E_{\lambda}(\vec{x}) = i \sum_j \sqrt{\frac{\omega_j}{2V \epsilon_0}} \left( a_{\lambda j} \mathrm{e}^{-i(\omega_j t - \vec{k}_j\cdot \vec{x})} -a^{\dag}_{\lambda j} \mathrm{e}^{i(\omega_j t - \vec{k}_j\cdot \vec{x})} \right) ,
 \end{equation}
 \begin{equation}
 q_n (\vec{x}) = \frac{1}{\sqrt{2 m_n \Omega_n V_S}} \left( b_n^\dag \mathrm{e}^{i(\Omega_n t - \vec{u}_n\cdot \vec{x})} + b_n \mathrm{e}^{-i(\Omega_n t - \vec{u}_n\cdot \vec{x})}  \right) ,
 \end{equation}
 \begin{equation}
 p_n (\vec{x}) = i \sqrt{\frac{m_n \Omega_n}{2 V_S}} \left( b_n^\dag \mathrm{e}^{i(\Omega_n t - \vec{u}_n\cdot \vec{x})} - b_n \mathrm{e}^{-i(\Omega_n t - \vec{u}_n\cdot \vec{x})}  \right) .
 \end{equation}
 In writing the previous expressions we used $\vec{u}_n, \vec{k}_j$ for the vibration and electric field wave vectors, respectively, $\Omega_n, \omega_j$ are the frequencies of the lattice vibration and electric field, $m_n$ is the effective mass of the $n$th normal mode, $V$ is the quantization volume of the electric field and $V_S$ is the volume of the sample. 
 The interaction Hamiltonian without any approximation is therefore obtained integrating $U^{\mathrm{int}}(\vec{x})$ over the volume of the sample and reads 
 \begin{align}
 \mathcal{H}_{\mathrm{int}} =& \int_{V_S} \mathrm{d}\vec{x} \epsilon_0 \sum_{\lambda \lambda'}\left[ \chi^{(0)}_{\lambda \lambda'} + \sum_n \chi^{(1)}_{n; \lambda \lambda'} \frac{1}{\sqrt{2 m_n \Omega_n V_S}} \left( b_n^\dag \mathrm{e}^{i(\Omega_n t - \vec{u}_n\cdot \vec{x})} + b_n \mathrm{e}^{-i(\Omega_n t - \vec{u}_n\cdot \vec{x})}  \right)  \right] \times \\
 &\times \sum_{j \ell} \frac{\sqrt{\omega_j \omega_\ell}}{2V\epsilon_0} \left( a_{\lambda j} \mathrm{e}^{-i(\omega_j t - \vec{k}_j\cdot \vec{x})} -a^{\dag}_{\lambda j} \mathrm{e}^{i(\omega_j t - \vec{k}_j\cdot \vec{x})}  \right)\left( a_{\lambda' \ell} \mathrm{e}^{-i(\omega_\ell t - \vec{k}_\ell\cdot \vec{x})} -a^{\dag}_{\lambda' \ell} \mathrm{e}^{i(\omega_\ell t - \vec{k}_\ell\cdot \vec{x})}  \right).
 \end{align}
Assuming periodic boundary conditions, the integration of the $\vec{x}$-dependent terms,  gives Kronecker deltas describing conservation of momentum
 \begin{equation}
 \int_{V_S} \mathrm{d}\vec{x} \,\mathrm{e}^{i(\vec{k}_j - \vec{k}_\ell  - \vec{u}_n)\cdot \vec{x}} = V_S \, \delta_{\vec{k}_j, \vec{k}_\ell + \vec{u}_n} .
 \end{equation}
 By concentrating on long wavelength phonons (with respect to the photon one), such that $u_n \simeq 0$, the problem becomes unidimensional because the wave-vector of the incident photon coincides with the wave-vector of the emitted photon. We can label the axis containing all the wave-vectors as $z$. 
Finally, the experimental evidence suggests that we can discard the terms like $aa$ or $a^\dag a^\dag$ and concentrate on those terms $a^\dag_{\lambda j} a_{\lambda' \ell}$ where $\omega_j-\omega_\ell = \Omega_n$ or $\omega_j-\omega_\ell = 0$.
In the end, we get an equilibrium rotation term,
\begin{equation}
H_{Ref}^{(0)}= -\frac{\omega_0 V_S}{2V}  \sum_{\lambda, \lambda'; j} \chi^{(0)}_{\lambda \lambda'} \Big( a^\dag_{\lambda j} a_{\lambda' j} + a^\dag_{\lambda' j} a_{\lambda j}  \Big), 
\end{equation}
a phonon-dependent modulation of this equilibrium rotation, where for simplicity we substitute the operator $(b_n+b_n^\dag)$ with its mean value with respect to the initial state
\begin{equation}
H_{Ref}^{(1)} = 2 \sum_{\lambda\lambda', j,n}  \, \widetilde{\chi}^{(1)}_{\lambda \lambda'}  a^\dag_{\lambda j} a_{\lambda' j}  \braket{b_n+b_n^\dag},
\end{equation}
and the Raman Hamiltonian
\begin{equation}
\label{Ram1}
H_{Ram}= \sum_{\lambda \lambda', j, n}  \, \widetilde{\chi}^{(1)}_{\lambda \lambda',n}\Big[\Big(\,a^\dag_{\lambda j}a_{\lambda' j+\frac{\Omega}{\delta}}\Big)\,b_n^\dag\,+\,\Big(\,a_{\lambda j}a^\dag_{\lambda' j+\frac{\Omega}{\delta}}\Big)\,b_n\Big]\ , \quad \widetilde{\chi}^{(1)}_{\lambda \lambda',n} = -\frac{\omega_0\sqrt{V_S}}{2V\sqrt{ 2m \Omega}} \chi^{(1)}_{\lambda \lambda',n}.
\end{equation}
Since the coefficients in the equilibrium term are such that $\chi^{(0)}_{\lambda \lambda'}= \chi^{(0) *}_{\lambda' \lambda}$, the Hamiltonian $H_{Ref}^{(0)}$ can be conveniently rewritten in the form
\begin{equation}
H_{Ref}^{(0)}=  \sum_{\lambda, \lambda'; j} \widetilde{\chi}^{(0)}_{\lambda \lambda'} a^\dag_{\lambda j} a_{\lambda' j}, \quad \widetilde{\chi}^{(0)}_{\lambda \lambda'} = -\frac{\omega_0 V_S}{V}Re\left(\chi^{(0)}_{\lambda \lambda'}\right),
\end{equation}
For simplicity, in writing \eqref{Ram1}, we used the approximation $\sqrt{\omega_j \omega_k}\sim \omega_0$, which is justified because in the following we are considering photonic frequencies in the range $350 \,\mathrm{THz}< \omega_j< 410 \,\mathrm{THz}$ with $\omega_0 = 380 \, \mathrm{THz}$. We shall collect the coefficients $\widetilde{\chi}^{(0)}_{\lambda \lambda'}$ and $\widetilde{\chi}^{(1)}_{\lambda \lambda'}$ (for each phonon mode $n$) into $2\times 2$ hermitian matrices $\boldsymbol{\widetilde{\chi}}^{(0)}$ and $\boldsymbol{\widetilde{\chi}}^{(1)}$ respectively.

\subsection{Dynamics}

The photon-phonon initial state evolves according to the following evolution operator during the time $\tau$ in which the photon crosses the sample 
\begin{equation}
U_{Bulk} = \mathrm{e}^{-i \tau ( H_{Ref}^{(0)} + H_{Ref}^{(1)}  + H_{Ram}) } = \mathrm{e}^{-i\tau H_{Ref}^{(0)} } \Big( 1 - i\int_0^\tau \mathrm{d}s \, (H_{Ref}^{(1)}(s) + H_{Ram}(s) ) + \ldots \Big)
\end{equation}
where $H_{Ref}^{(1)}(s) = \mathrm{e}^{i s H_{Ref}^{(0)}} \, H_{Ref}^{(1)} \, \mathrm{e}^{-i s H_{Ref}^{(0)}} $ and $H_{Ram}(s)= \mathrm{e}^{i s H_{Ref}^{(0)}} H_{Ram} \mathrm{e}^{-i s H_{Ref}^{(0)}} $. The following structure is assumed for the matrix $\boldsymbol{\widetilde{\chi}}^{(0)}$ 
\begin{equation}
\boldsymbol{\widetilde{\chi}}^{(0)} = 
 \begin{pmatrix}
       u  & \overline{w} \\
       \overline{w}  & u
       \end{pmatrix},
\end{equation} 
where $\overline{w}= w\cos(\phi)$, being $\phi$ the phase of the complex coefficient $\chi^{(0)}_{x y}$.
The experimental evidence is that, when a linearly polarized light beam, say along the $x$ axis, interacts with the bulk of the sample, the polarization of the transmitted light becomes elliptic and the main axis of the ellipsis is rotated with respect to $x$. The non-vanishing off-diagonal elements in $\boldsymbol{\widetilde{\chi}}^{(0)}$ account for the rotation, while the ellipticity is given by the phase $\phi$. In order to compensate for this rotation (but not for ellipticity), one then rotates the polarization analyser along the directions $x'$ and $y'$, therefore effectively measuring the intensities $I_{x'}=U_{Rot}(\tau) \, a^\dag_{x j} a_{x j} \, U_{Rot}^{\dag}(\tau) $ and $ I_{y'} = U_{Rot}(\tau) \, a^\dag_{y j} a_{y j} \, U_{Rot}^{\dag}(\tau) $, where the unitary evolution operator $U_{Rot}(\tau)= \mathrm{e}^{-i \tau H^{Rot}}$, with
\begin{equation}
H^{Rot}= \sum_{\lambda \lambda'}\chi^{Rot}_{\lambda \lambda'} \, a^\dag_{\lambda j} a_{\lambda' j} , \quad 
\boldsymbol{\chi}^{Rot}= \begin{pmatrix}
       u  & w \\
       w  & u
       \end{pmatrix} ,
\end{equation}
describes the action of the analyser using a rotation matrix that is analogous to $\boldsymbol{\widetilde{\chi}}^{(0)}$ but without the phase.

Considering an initial coherent state $|\alpha\rangle \langle \alpha|$ for the light pulse and a state $\varrho_t$ for the vibrational degree of freedom, the calculation to be performed reads
\begin{equation}
\braket{I_{\lambda' j}(\tau)} = \mathrm{Tr}[\varrho_t(\tau) I_{\lambda' j} ] = \Big[  \big(|\alpha\rangle \langle \alpha| \otimes \varrho_t \big) \,U_{Bulk}^\dag (\tau) \, U_{Rot}(\tau) \, a^\dag_{\lambda j} a_{\lambda j} \, U_{Rot}^{\dag}(\tau)\, U_{Bulk}(\tau) \Big].
\end{equation}
Since the dynamics induced by the atomic displacement can be considered a perturbation with respect to the equilibrium rotation, one can conveniently treat both the phonon-dependent rotation and the Raman contribution using perturbation theory. In particular, one introduces the map $\mathcal{L}$ as follows
\begin{equation}
\mathcal{L}[a^\dag_{\lambda j} a_{\lambda j}]=  \mathrm{e}^{i \tau H_{Ref}^{(0)}} \mathrm{e}^{-i \tau H_{Rot}} a^\dag_{\lambda j} a_{\lambda j} \mathrm{e}^{i \tau H_{Rot}} \mathrm{e}^{-i \tau H_{Ref}^{(0)}}
\end{equation}
and use it to rewrite the dynamical evolution in a convenient way
\begin{equation}\label{approx}
U_{Bulk}^\dag (\tau) \, U_{Rot}(\tau) \, a^\dag_{\lambda j} a_{\lambda j} \, U_{Rot}^{\dag}(\tau)\, U_{Bulk}(\tau) \simeq \mathcal{L}[a^\dag_{\lambda j} a_{\lambda j}] + i \int_{0}^{\tau}\mathrm{d}s \Big[ H_{Ref}^{(1)}(s) + H_{Ram}(s), \mathcal{L}[a^\dag_{\lambda j} a_{\lambda j}] \Big].
\end{equation}
A key element for the following calculation is the action of the operator $U_{Rot}^{(0)}$ on the annihilation operator $a_{\mu k}$
\begin{equation}\label{rot}
U_{Ref}^{(0) \dag}(s) \, a_{\mu k} \, U_{Ref}^{(0)}(s) = \sum_{\lambda} \big( C_{\mu \lambda}(s) - i S_{\mu \lambda}(s) \big) a_{\lambda k} = \sum_{\lambda} R_{\mu \lambda}(s) a_{\lambda k} 
\end{equation}
where $C_{\mu \lambda}(s) = \Big( \cos(s \boldsymbol{\tilde{\chi}}^{(0)} ) \Big)_{\mu \lambda}$ and $S_{\mu \lambda}(s) = \Big( \sin(s \boldsymbol{\tilde{\chi}}^{(0)} ) \Big)_{\mu \lambda}$. These coefficients can be given explicitly using the structure of the matrix $\boldsymbol{\tilde{\chi}}^{(0)}$, namely
\begin{align*}
&\boldsymbol{\widetilde{{\chi}}}^{(0)}= 
\begin{pmatrix}
u &\overline{w} \\
\overline{w} &u 
\end{pmatrix}, \quad 
\cos(\boldsymbol{\widetilde{{\chi}}}^{(0)})= 
\begin{pmatrix}
\cos(u)\cos(\overline{w}) &-\sin(u)\sin(\overline{w}) \\
-\sin(u)\sin(\overline{w}) & \cos(u)\cos(\overline{w})
\end{pmatrix}, \\
&\sin(\boldsymbol{\widetilde{{\chi}}}^{(0)})= 
\begin{pmatrix}
\sin(u)\cos(\overline{w}) &\cos(u)\sin(\overline{w}) \\
\cos(u)\sin(\overline{w}) & \sin(u)\cos(\overline{w})
\end{pmatrix}.
\end{align*}
Using \eqref{rot} one can write the action of $\mathcal{L}$ on the operators $a^\dag_{y k} a_{y k}$ and $a^\dag_{x k} a_{x k}$ as
\begin{equation}\label{Ly}
\mathcal{L}[a^\dag_{y k} a_{y k}]= \sum_{\mu \mu'}\sum_{\lambda \lambda'} R^{rot *}_{y \mu}(-\tau) R^{rot}_{y \mu'}(-\tau) R^*_{\mu \lambda}(\tau) R_{\mu' \lambda'}(\tau) a^\dag_{\lambda j} a_{\lambda' j},
\end{equation}
\begin{equation}\label{Lx}
\mathcal{L}[a^\dag_{x k} a_{x k}]= \sum_{\mu \mu'}\sum_{\lambda \lambda'} R^{rot *}_{x \mu}(-\tau) R^{rot}_{x \mu'}(-\tau) R^*_{\mu \lambda}(\tau) R_{\mu' \lambda'}(\tau) a^\dag_{\lambda j} a_{\lambda' j}.
\end{equation}
Also, one could write the action of $\mathcal{L}$ implicitly as
\begin{equation}
\mathcal{L}[a^\dag_{y k} a_{y k}] = A^y a^\dag_{y k} a_{y k} + B^y a^\dag_{x k} a_{x k} + D^y a^\dag_{x k} a_{y k} + D^{y *} a^\dag_{y k} a_{x k},
\end{equation}
\begin{equation}
\mathcal{L}[a^\dag_{x k} a_{x k}] = A^x a^\dag_{y k} a_{y k} + B^x a^\dag_{x k} a_{x k} + D^x a^\dag_{x k} a_{y k} + D^{x *} a^\dag_{y k} a_{x k},
\end{equation}
where the coefficients $A^\lambda, B^\lambda$ and $D^\lambda$, with $\lambda\in\{x,y\}$ are to be calculated. 
By comparison, the coefficients can be calculated as follows
\begin{align}
A^y&= \sum_{\mu \mu'}  R^{rot *}_{y \mu}(-\tau) R^{rot}_{y \mu'}(-\tau) R^*_{\mu y}(\tau) R_{\mu' y}(\tau) \\
B^y&= \sum_{\mu \mu'}  R^{rot *}_{y \mu}(-\tau) R^{rot}_{y \mu'}(-\tau) R^*_{\mu x}(\tau) R_{\mu' x}(\tau) \\
D^y&= \sum_{\mu \mu'}  R^{rot *}_{y \mu}(-\tau) R^{rot}_{y \mu'}(-\tau) R^*_{\mu x}(\tau) R_{\mu' y}(\tau) 
\end{align}
\begin{align}
A^x&= \sum_{\mu \mu'}  R^{rot *}_{x \mu}(-\tau) R^{rot}_{x \mu'}(-\tau) R^*_{\mu y}(\tau) R_{\mu' y}(\tau) \\
B^x&= \sum_{\mu \mu'}  R^{rot *}_{x \mu}(-\tau) R^{rot}_{x \mu'}(-\tau) R^*_{\mu x}(\tau) R_{\mu' x}(\tau) \\
D^x&= \sum_{\mu \mu'}  R^{rot *}_{x \mu}(-\tau) R^{rot}_{x \mu'}(-\tau) R^*_{\mu x}(\tau) R_{\mu' y}(\tau) 
\end{align}
and explicitly they read
\begin{align}
A^y&= B^x = \cos^2(\tau(\overline{w}-w))  \\
B^y&= A^x = \sin^2(\tau(\overline{w}-w))\\
D^y&=- D^x = \frac{i}{2} \sin(2\tau(\overline{w}-w))
\end{align}

\subsubsection{LRM}
One can then evaluate the commutator of the dressed Hamiltonian $H_{Ref}^{(1)}(s)$, given by the following expression (we dropped the phonon dependence $n$ because we are considering a single mode)
\begin{equation}
H_{Ref}^{(1)}(s) = 2\sum_{\lambda \lambda'; j} \tilde{\chi}^{(1)}_{\lambda \lambda'}\Big( \sum_{\mu \mu'} R^*_{\lambda \mu}(s) R_{\lambda' \mu'}(s) a^\dag_{\mu j} a_{\mu' j} \Big) \braket{b+b^\dag},
\end{equation}
with each term $a^\dag_{\eta k} a_{\eta' k}$ in \eqref{Lx} and \eqref{Ly}
\begin{equation}
\Big[ H_{Ref}^{(1)}(s), a^\dag_{\eta k} a_{\eta' k} \Big] = 2 \sum_{\lambda \lambda'}  \tilde{\chi}^{(1)}_{\lambda \lambda'} \braket{b+b^\dag} \sum_{\mu}\Big( R^*_{\lambda \mu}(s) R_{\lambda' \eta}(s) a^{\dag}_{\mu k} a_{\eta' k} - R^*_{\lambda \eta'}(s) R_{\lambda' \mu}(s) a^\dag_{\eta k} a_{\mu k}  \Big).
\end{equation}
Since the impinging light beam is linearly polarized along the $x$ axis, the relevant terms in the previous expression are those proportional to $a^\dag_{x k} a_{x k}$. Therefore, when considering the average with respect to the state $|\alpha\rangle \langle \alpha| \otimes \varrho_t $ the non vanishing terms have $\mu=x$
\begin{equation}
\braket{\Big[ H_{Ref}^{(1)}(s), a^\dag_{\eta k} a_{\eta' k} \Big]} = 2 \sum_{\lambda \lambda'}  \tilde{\chi}^{(1)}_{\lambda \lambda'} \braket{b+b^\dag} \Big( R^*_{\lambda x}(s) R_{\lambda' \eta}(s) \braket{a^{\dag}_{x k} a_{\eta' k}} - R^*_{\lambda \eta'}(s) R_{\lambda' x}(s) \braket{a^\dag_{\eta k} a_{x k}}  \Big).
\end{equation}
Moreover, the terms with $\eta=\eta'$ are both vanishing and the only two contributions come from
\begin{enumerate}
\item $\eta=x, \eta'=y$
\begin{equation}\label{ref1}
\braket{\Big[ H_{Ref}^{(1)}(s), a^\dag_{x k} a_{y k} \Big]} = - 2\sum_{\lambda \lambda'} \tilde{\chi}^{(1)}_{\lambda \lambda'} R^*_{\lambda y}(s) R_{\lambda' x}(s) \braket{b+b^\dag} |\alpha_k|^2 ,
\end{equation}
\item $\eta=y, \eta'=x$
\begin{equation}\label{ref2}
\braket{\Big[ H_{Ref}^{(1)}(s), a^\dag_{y k} a_{x k} \Big]} = 2\sum_{\lambda \lambda'} \tilde{\chi}^{(1)}_{\lambda \lambda'} R^*_{\lambda' x}(s) R_{\lambda y}(s) \braket{b+b^\dag} |\alpha_k|^2 .
\end{equation}
\end{enumerate}

\subsubsection{Raman}

One has to compute also the Raman contribution to the modulation of the intensity. The Raman dressed Hamiltonian reads
\begin{equation}
H_{Ram}(s)= \mathrm{e}^{i s H_{Ref}^{(0)}} H_{Ram} \mathrm{e}^{-i s H_{Ref}^{(0)}} = \sum_{\lambda \lambda'} \tilde{\chi}^{(1)}_{\lambda \lambda'} \sum_{\mu \mu'} \Big( R^*_{\lambda \mu}(s) R_{\lambda' \mu'}(s) a^\dag_{\mu j} a_{\mu' j+\frac{\Omega}{\delta}} b^\dag + R^*_{\lambda' \mu'}(s) R_{\lambda \mu}(s) a^\dag_{\mu' j+\frac{\Omega}{\delta}} a_{\mu j} b  \Big) 
\end{equation}
Consider the commutator 
\begin{align}
\Big[ H_{Ram}^{(1)}(s), a^\dag_{\eta k} a_{\eta' k} \Big] = &\sum_{\lambda \lambda'}\tilde{\chi}^{(1)}_{\lambda \lambda'} \sum_{\mu} \Big[ \Big( R^*_{\lambda \mu}(s) R_{\lambda' \eta}(s) a^\dag_{\mu k-\frac{\Omega}{\delta}} a_{\eta' k} - R^*_{\lambda \eta'}(s) R_{\lambda' \mu}(s) a^\dag_{\eta k} a_{\mu k+\frac{\Omega}{\delta}} \Big) b^\dag +  \nonumber \\ 
& +\Big(  R^*_{\lambda' \mu}(s) R_{\lambda \eta}(s) a^\dag_{\mu k+\frac{\Omega}{\delta}} a_{\eta' k}  - R^*_{\lambda' \eta'}(s) R_{\lambda \mu}(s) a^\dag_{\eta k} a_{\mu k-\frac{\Omega}{\delta}} \Big)b \Big].
\end{align}
By taking its mean value with respect to the photon state $\vert\alpha\rangle$ in \eqref{coherent}, only the terms with both operators related to the $x$ polarization contribute. In particular, one has three cases
\begin{enumerate}
\item $\eta = x$ and $\eta'=y$
\begin{equation}
\braket{\Big[ H_{Ram}^{(1)}(s), a^\dag_{x k} a_{y k} \Big] } = - \sum_{\lambda \lambda'}\tilde{\chi}^{(1)}_{\lambda \lambda'} |\alpha_k| \Big(|\alpha_{k+\frac{\Omega}{\delta}}| \braket{b^\dag} + |\alpha_{k-\frac{\Omega}{\delta}}| \braket{b}  \Big) R^*_{\lambda y}(s) R_{\lambda' x}(s)
\end{equation}
\item $\eta = y$ and $\eta'=x$
\begin{equation}
\braket{\Big[ H_{Ram}^{(1)}(s), a^\dag_{y k} a_{x k} \Big] } = \sum_{\lambda \lambda'}\tilde{\chi}^{(1)}_{\lambda \lambda'} |\alpha_k| \Big(|\alpha_{k-\frac{\Omega}{\delta}}| \braket{b^\dag} + |\alpha_{k+\frac{\Omega}{\delta}}| \braket{b}  \Big) R_{\lambda y}(s) R^*_{\lambda' x}(s)
\end{equation}
\item $\eta= x$ and $\eta' =x$
\begin{equation}
\braket{\Big[ H_{Ram}^{(1)}(s), a^\dag_{x k} a_{x k} \Big] } = \sum_{\lambda \lambda'}\tilde{\chi}^{(1)}_{\lambda \lambda'} |\alpha_k| \Big(|\alpha_{k-\frac{\Omega}{\delta}}|- |\alpha_{k+\frac{\Omega}{\delta}}| \Big) \Big(\braket{b^\dag} - \braket{b} \Big) R^*_{\lambda x}(s) R_{\lambda' x}(s)
\end{equation}
\end{enumerate}


In the main text the action of the Raman Hamiltonian is described also on its own, without the rotation provided by $H^{(0)}_{Ref}$, in order to highlight its main features. We also use a weaker approximation on the factor $\sqrt{\omega_j(\omega_j \pm\Omega)}\sim \omega_j$.
The time-evolution of an operator $O$ under the action of the unitary propagator $U_{Ram}(\tau)=\mathrm{e}^{-i \tau H_{Ram}}$ is given in this case by
\begin{equation}
O(\tau)= U_{Ram}^\dag (\tau) \, O \, U_{Ram}(\tau),
\end{equation}
and up to second order in the Raman coupling one can write
\begin{equation} \label{approx}
O(\tau) \simeq O +i\tau \Big[H_{Ram}, O \Big]-\frac{\tau^2}{2}\Big[H_{Ram},\Big[H_{Ram}, O \Big]\Big] . 
\end{equation}
The time-evolution of specific observables induced by the Raman Hamiltonian can be readily studied. The first example is the phononic operator $b$.
For this operator the second order term, {\it i.e.} the double commutator in \eqref{approx}, turns out to be zero and the time-evolution reads
\begin{equation}\label{b}
b(\tau) \simeq  b+ i\tau \frac{\sqrt{V_S}}{2V\sqrt{ 2m \Omega}} g , \quad g= \sum_{\lambda \lambda', j}\chi^{(1)}_{\lambda \lambda'}\, \omega_j\, a^\dag_{\lambda j} a_{\lambda' j+\frac{\Omega}{\delta}} ,
\end{equation}
where the operator $g$ has been introduced for future convenience. 
By taking the average with respect to a factorized state $|\alpha\rangle \langle \alpha| \otimes \varrho$, where $\varrho$ is a generic density matrix for the phononic degrees of freedom and $|\alpha\rangle$ is a coherent state as described in \eqref{coherent} and \eqref{amplitudes}, one obtains
\begin{equation}
\braket{b(\tau)}:= \mathrm{Tr}\Big[ \big(|\alpha\rangle \langle \alpha| \otimes \varrho \big)\, b(\tau)  \Big] = \braket{b}  +i\tau \frac{\sqrt{V_S}}{2V\sqrt{ 2m \Omega}}\gamma , \quad   \gamma= \braket{g} = \sum_{\lambda \lambda', j} \chi^{(1)}_{\lambda \lambda'}\, \omega_j \, |\alpha_{\lambda j}| |\alpha_{\lambda' j+\frac{\Omega}{\delta}}|.
\end{equation}
Using the previous result one can compute the mean phonon position which is proportional to $b+b^\dag$ and discover that up to second order in the Raman coupling this is not modified 
\begin{equation}
\braket{q(\tau)}= \frac{1}{\sqrt{2 m \Omega V_S}} \Big(\braket{b(\tau)} + \braket{b^\dag(\tau)} \Big) = \braket{q(0)} ,
\end{equation}
The mean phonon momentum is instead shifted by an amount proportional to the interaction time $\tau$ 
\begin{align}
\braket{p(\tau)}=i \sqrt{\frac{m\Omega}{2V_S}}\Big(\braket{b^\dag(\tau)} - \braket{b(\tau)}\Big) = \braket{p(0)}+ \frac{\tau}{2V}|\gamma| .
\end{align} 
Analogously, for the phonon number operator one can explicitly compute
\begin{equation}
N(\tau)=( b^\dag  b ) (\tau) \simeq b^\dag b + i\tau \sum_{\lambda \lambda', j}\widetilde{\chi}^{(1)}_{\lambda \lambda'}\omega_j \left(  a^\dag_{\lambda' j+\frac{\Omega}{\delta}} a_{\lambda j} b - a^\dag_{\lambda j} a_{\lambda' j+\frac{\Omega}{\delta}} b^\dag  \right) +\tau^2 \frac{V_S}{8V^2 m \Omega}  g^\dag g  ,
\end{equation}
and the mean number of phonons after the Raman interaction is
\begin{equation}
\braket{N(\tau)}= \braket{N(0)} +  \frac{\tau V_S |\gamma|}{2Vm\Omega} \braket{p(0)} + \tau^2 \frac{V_S}{8V^2 m \Omega} \braket{g^\dag g}.
\end{equation}

Concerning the photonic operators $a_{\mu j}$ the Raman evolution yields
\begin{equation*}
a_{\mu j}(\tau) \simeq a_{\mu j} +i\tau \Big[H_{Ram}, a_{\mu j}\Big]-\frac{\tau^2}{2}\Big[H_{Ram},\Big[H_{Ram}, a_{\mu j}\Big]\Big] + \dots
\end{equation*}
where the first order contribution reads
\begin{equation*}
\Big[H_{Ram}, a_{\mu j}\Big] = - \sum_{\lambda}\widetilde{\chi}^{(1)}_{\mu \lambda} \omega_j \left(  a_{\lambda j+\frac{\Omega}{\delta}} \, b^\dag + a_{\lambda j-\frac{\Omega}{\delta}} \, b \right)
\end{equation*}
while the second order one has the form:
\begin{align*}
\Big[H_{Ram},\Big[H_{Ram}, a_{\mu j}\Big]\Big] &= \sum_{\lambda \eta} \widetilde{\chi}^{(1)}_{\mu \lambda}\, \widetilde{\chi}^{(1)}_{\lambda \eta} \omega_j^2 \Big(   a_{\eta j+\frac{2\Omega}{\delta}} \, \left( b^\dag \right)^2 +  a_{\eta j}\, b b^\dag +  a_{\eta j}\, b^\dag b +   a_{\eta j-\frac{2\Omega}{\delta}}\, b^2 \Big)+ \\
&\quad +\sum_{\lambda \eta \eta' k} \widetilde{\chi}^{(1)}_{\mu \lambda}\, \widetilde{\chi}^{(1)}_{\eta \eta'}\omega_k \omega_j \Big(  a^\dag_{\eta k} \, a_{\eta' k+\frac{\Omega}{\delta}}\, a_{\lambda j-\frac{\Omega}{\delta}} -  a^\dag_{\eta' k+\frac{\Omega}{\delta}} \, a_{\eta k} \, a_{\lambda j+\frac{\Omega}{\delta}} \Big) .
\end{align*}
The average transmitted intensity $I_{\lambda j}=a^\dag_{\lambda j}a_{\lambda j}$ therefore reads
\begin{equation}
\braket{I_{\lambda j}(\tau)} 
=\braket{I_{\lambda j}(0)}+ \tau \frac{V_S}{2Vm\Omega} \sum_{\lambda’} \chi^{(1)}_{\lambda\lambda’}\omega_j |\alpha_{\lambda j}|\Big(|\alpha_{\lambda’ j+\frac{\Omega}{\delta}}|-|\alpha_{\lambda’ j-\frac{\Omega}{\delta}}| \Big)\left(\braket{p(0)} +\frac{\tau \gamma}{4V} \right) +\tau^2 \gamma'_j,
\end{equation}
where the momentum dependence of the first order term has been highlighted, as well as a second order term with a similar structure apart from a parameter $\gamma$ replacing $p$ and a further second order term, indicated by $\gamma'_j$, containing contributions of second order in the phononic operators. Explicitly, 
\begin{align}
\gamma'_j =& \sum_{\lambda \eta} \widetilde{\chi}^{(1)}_{\mu \lambda}\, \widetilde{\chi}^{(1)}_{\mu \eta} \omega_j^2 \Big( |\alpha_{\lambda j+\frac{\Omega}{\delta}}| |\alpha_{\eta j+\frac{\Omega}{\delta}}| \braket{b b^\dag} + |\alpha_{\lambda j-\frac{\Omega}{\delta}}| |\alpha_{\eta j-\frac{\Omega}{\delta}}| \braket{b^\dag b} + |\alpha_{\lambda j-\frac{\Omega}{\delta}}| |\alpha_{\eta j+\frac{\Omega}{\delta}}| \Big(  \braket{b^{\dag 2}} +  \braket{b^2} \Big) \Big) \nonumber \\
&-\frac{1}{2} \sum_{\lambda \eta} \widetilde{\chi}^{(1)}_{\mu \lambda}\, \widetilde{\chi}^{(1)}_{\lambda \eta} \omega_j^2 |\alpha_{\mu j}| \Big( \left( |\alpha_{\eta j+\frac{2\Omega}{\delta}}| + |\alpha_{\eta j-\frac{2\Omega}{\delta}}| \right) \Big(  \braket{b^{\dag 2}} + \braket{b^2} \Big) + 2 |\alpha_{\eta j}| \braket{b b^\dag + b^\dag b}  \Big) .
\end{align}

In the following, we keep discussing the complete model (Raman $+$ LRM) introduced at the beginning.

\subsubsection{Time-dependence}
The dependence on the delay time between pump and probe is encoded in the average of the phononic operators $\braket{b}$. Indeed, we compute $\braket{b}$ as follows
\begin{equation}
\braket{b} = \mathrm{Tr}\Big[|\alpha^{pump}\rangle \langle \alpha^{pump}|\otimes \varrho_\beta \, U^\dag_{Bulk}(\tau) U^\dag_{free}(t) \, b \, U_{free}(t) U_{Bulk}(\tau)\Big],
\end{equation}
where $U_{free}(t)= \mathrm{e}^{-i t \Omega b^\dag b}$, assuming the vibration in the crystal evolving without dissipation in the time interval $t$. In order to make explicit the dependence of its action on the pump polarization angle with respect to the $x$ axis, we choose the coherent initial state of the pump $| \alpha^{pump} \rangle$ such that $a_{x j} | \alpha^{pump} \rangle= \alpha^{pump}_{x j} | \alpha^{pump} \rangle $ and $a_{y j} | \alpha^{pump} \rangle= \alpha^{pump}_{y j} | \alpha^{pump} \rangle $ where
\begin{equation}
\alpha^{pump}_{x j}= \alpha^{pump}_{j} \cos(\theta)  , \quad \alpha^{pump}_{y j}= \alpha^{pump}_{j} \sin(\theta).
\end{equation}
Moreover, we assume the state of the vibrational degree of freedom before the action of the pump to be a thermal state $\varrho_\beta= \frac{\mathrm{e}^{-\beta \Omega b^\dag b}}{\mathrm{Tr}[\mathrm{e}^{-\beta \Omega b^\dag b}]}$.

The computation can be done perturbatively 
\begin{equation}
U^\dag_{Bulk}(\tau) U^\dag_{free}(t) \, b \, U_{free}(t) U_{Bulk}(\tau)\simeq \mathrm{e}^{-i\Omega t}\Big( b+i\int_0^\tau \mathrm{d}s \Big[ H^{(1)}_{Ref}(s) + H_{Ram}(s), b \Big] \Big).
\end{equation}
The term with $\mathrm{Tr}[\varrho_\beta b]$ gives no contribution because $\varrho_\beta$ is a thermal state. The commutators can be computed using the expression for $H^{(1)}_{Ref}(s)$ and $H_{Ram}(s)$ that we already found. In particular,
\begin{equation}
\Big[ H^{(1)}_{Ref}(s) , b \Big]= 0
\end{equation}
because of the parametric dependence, while
\begin{equation}
\Big[ H_{Ram}(s) , b \Big]= -\sum_{\lambda \lambda' j} \tilde{\chi}^{(1)}_{\lambda \lambda'}\sum_{\mu \mu'} R^*_{\lambda \mu}(s) R_{\lambda' \mu'}(s) a^\dag_{\mu j} a_{\mu' j+\frac{\Omega}{\delta}}.
\end{equation}
Moreover, using the explicit form of the state $|\alpha^{pump}\rangle \langle \alpha^{pump}|$, the quantum expectation value reads
\begin{equation}
\braket{\Big[ H_{Ram}(s) , b \Big]} = - \sum_j |\alpha^{pump}_j| |\alpha^{pump}_{j+\frac{\Omega}{\delta}}| M(s)
\end{equation}
where the factor $M(s)$ is given by
\begin{equation}\label{ms}
M(s)= \sum_{\lambda \lambda'} \tilde{\chi}^{(1)}_{\lambda \lambda'} \Big[ R^*_{\lambda x}(s) R_{\lambda' x}(s)\cos^2(\theta) + R^*_{\lambda y}(s) R_{\lambda' y}(s)\sin^2(\theta) + \Big(R^*_{\lambda x}(s) R_{\lambda' y}(s) + R^*_{\lambda y}(s) R_{\lambda' x}(s)\Big)\sin(\theta)\cos(\theta) \Big].
\end{equation}
More explicit expression will be given in the following when discussing the different vibrational mode of quartz, thus providing the structure of the matrix $\boldsymbol{\tilde{\chi}}^{(1)}$.

\section{Quartz}

Three different phonons of quartz are studied with our setup, indicated in brief as $A$, $E_L$ and $E_T$. The matrix $\boldsymbol{\tilde{\chi}}^{(1)}$ has therefore a different structure depending on the specific phonon excited. In particular, one has
\begin{align}
A: \quad  
&\boldsymbol{\tilde{\chi}}^{(1)}= 
\begin{pmatrix}
a  &0 \\
0   & a
\end{pmatrix},     \\
E_L: \quad
&\boldsymbol{\tilde{\chi}}^{(1)}= 
\begin{pmatrix}
c_L  &0 \\
0   & -c_L
\end{pmatrix},     \\
E_T: \quad
&\boldsymbol{\tilde{\chi}}^{(1)}= 
\begin{pmatrix}
0  &-c_T \\
-c_T   & 0
\end{pmatrix}.    
\end{align}
Since our model treats perturbatively the phonon-dependent modulation up to first order, the three modes can be studied separately. The final result is then obtained by summing up the three different contributions.

\subsubsection{LRM}
First, we can compute the LRM according to \eqref{ref1} and \eqref{ref2} for each mode.
For the $A$ mode the $H^{(1)}_{Ref}$ commutes with the $H^{(0)}_{Ref}$ so that 
\begin{equation}
\braket{\Big[ H_{Ref}^{(1)}(s), a^\dag_{x k} a_{y k} \Big]} = \braket{\Big[ H_{Ref}^{(1)}, a^\dag_{x k} a_{y k} \Big]} = 0.
\end{equation}
The contribution related to $a^\dag_{y k} a_{x k}$ is also vanishing. For the longitudinal mode $E_L$ one finds
\begin{align}
\braket{\Big[ H_{Ref}^{(1)}(s), a^\dag_{x k} a_{y k} \Big]} &=- 2 c_L \braket{b+b^\dag}_{E_L} |\alpha_k|^2 (R^*_{x y}(s) R_{x x}(s)- R^*_{y y}(s) R_{y x}(s))= \nonumber \\
&=- 2 i  \sin(2\overline{w}s) \,c_L \braket{b+b^\dag}_{E_L} |\alpha_k|^2
\end{align}
\begin{align}
\braket{\Big[ H_{Ref}^{(1)}(s), a^\dag_{y k} a_{x k} \Big]} &= 2 c_L \braket{b+b^\dag}_{E_L} |\alpha_k|^2 (R_{x y}(s) R^*_{x x}(s)- R_{y y}(s) R^*_{y x}(s))= \nonumber \\
&=- 2 i \sin(2\overline{w}s) \,c_L \braket{b+b^\dag}_{E_L} |\alpha_k|^2.
\end{align}
The integrals in \eqref{approx} can be promptly performed
\begin{equation}
i \int_0^\tau \mathrm{d}s \braket{\Big[ H_{Ref}^{(1)}(s), a^\dag_{x k} a_{y k} \Big]} = 2 c_L \braket{b+b^\dag}_{E_L} |\alpha_k|^2 \frac{1- \cos(2\overline{w}\tau)}{2\overline{w}} = 2  c_L \braket{b+b^\dag}_{E_L} |\alpha_k|^2 \frac{\sin^2(\overline{w}\tau)}{\overline{w}},
\end{equation}
\begin{equation}
i \int_0^\tau \mathrm{d}s \braket{\Big[ H_{Ref}^{(1)}(s), a^\dag_{y k} a_{x k} \Big]} =  2  c_L \braket{b+b^\dag}_{E_L} |\alpha_k|^2 \frac{\sin^2(\overline{w}\tau)}{\overline{w}},
\end{equation}
and it turns out that the $E_L$ modulation is vanishing because $D^y=-D^{y *}=-D^x$.
Finally, for the transverse mode $E_T$ one finds
\begin{align}
\braket{\Big[ H_{Ref}^{(1)}(s), a^\dag_{x k} a_{y k} \Big]} &=2 c_T \braket{b+b^\dag}_{E_T} |\alpha_k|^2 (R^*_{x y}(s) R_{y x}(s)+ R^*_{y y}(s) R_{x x}(s))= \nonumber \\
&= 2 c_T \braket{b+b^\dag}_{E_T} |\alpha_k|^2
\end{align}
\begin{align}
\braket{\Big[ H_{Ref}^{(1)}(s), a^\dag_{y k} a_{x k} \Big]} &= - 2c_T \braket{b+b^\dag}_{E_T} |\alpha_k|^2 (R_{x y}(s) R^*_{y x}(s)+ R_{y y}(s) R^*_{x x}(s))= \nonumber \\
&= - 2 c_T \braket{b+b^\dag}_{E_T} |\alpha_k|^2
\end{align}
The integrals in this case read
\begin{equation}
i \int_0^\tau \mathrm{d}s \braket{\Big[ H_{Ref}^{(1)}(s), a^\dag_{x k} a_{y k} \Big]} = i 2 \tau c_T \braket{b+b^\dag}_{E_T} |\alpha_k|^2  
\end{equation}
\begin{equation}
i \int_0^\tau \mathrm{d}s \braket{\Big[ H_{Ref}^{(1)}(s), a^\dag_{y k} a_{x k} \Big]} = -i 2 \tau  c_T \braket{b+b^\dag}_{E_T} |\alpha_k|^2 
\end{equation}

The LRM modulation of $\braket{I_{y' k}(\tau)}$ therefore is only due to the $E_T$ mode and reads
\begin{align}
i \int_0^\tau \mathrm{d}s D^y \braket{\Big[ H_{Ref}^{(1)}(s), a^\dag_{x k} a_{y k} \Big]} + i \int_0^\tau \mathrm{d}s D^{y *} \braket{\Big[ H_{Ref}^{(1)}(s), a^\dag_{y k} a_{x k} \Big]}  = - 2c_T \tau \braket{b + b^\dag}_{E_T}|\alpha_k|^2 \sin(2\tau(\overline{w}-w)).
\end{align}
The LRM modulation of $\braket{I_{x' k}(\tau)}$ has an analogous expression with the opposite sign because $D^x=-D^y$.

Since $w\tau$ is a small parameter in our case, one can expand perturbatively the previous results in order to get simpler expressions to be compared with the experiment. In particular, the refractive modulation of both $\braket{I_{y' k}(\tau)}$ and $\braket{I_{x' k}(\tau)}$ is of first order in $w\tau$
\begin{equation}
i \int_0^\tau \mathrm{d}s D^y \braket{\Big[ H_{Ref}^{(1)}(s), a^\dag_{x k} a_{y k} \Big]} + i \int_0^\tau \mathrm{d}s D^{y *} \braket{\Big[ H_{Ref}^{(1)}(s), a^\dag_{y k} a_{x k} \Big]} \simeq -c_T \tau \,4 w\tau \,(\cos(\phi)-1) \braket{b + b^\dag}_{E_T}|\alpha_k|^2 
\end{equation}

\subsubsection{Raman}
Concerning Raman, we can evaluate the three contributions for each phonon.
For the mode $A$ we have that only the third term is non-vanishing 
\begin{equation}
\braket{\Big[ H_{Ram}^{(1)}(s), a^\dag_{x k} a_{x k} \Big] } = a |\alpha_k| \Big(|\alpha_{k-\frac{\Omega_A}{\delta}}|- |\alpha_{k+\frac{\Omega_A}{\delta}}| \Big) \braket{b^\dag - b}_A 
\end{equation}
For the mode $E_L$ one has instead
\begin{equation}
\braket{\Big[ H_{Ram}^{(1)}(s), a^\dag_{x k} a_{y k} \Big] } =- c_L |\alpha_k| \Big(|\alpha_{k+\frac{\Omega_E}{\delta}}| \braket{b^\dag}_{E_L} + |\alpha_{k-\frac{\Omega_{E}}{\delta}}| \braket{b}_{E_L}  \Big) i  \sin(2\overline{w}s),
\end{equation}
\begin{equation}
\braket{\Big[ H_{Ram}^{(1)}(s), a^\dag_{y k} a_{x k} \Big] } =- c_L |\alpha_k| \Big(|\alpha_{k-\frac{\Omega_E}{\delta}}| \braket{b^\dag}_{E_L} + |\alpha_{k+\frac{\Omega_E}{\delta}}| \braket{b}_{E_L}  \Big) i \sin(2\overline{w}s),
\end{equation}
\begin{equation}
\braket{\Big[ H_{Ram}^{(1)}(s), a^\dag_{x k} a_{x k} \Big] } = c_L |\alpha_k| \Big(|\alpha_{k-\frac{\Omega_E}{\delta}}|- |\alpha_{k+\frac{\Omega_E}{\delta}}| \Big) \braket{b^\dag - b}_{E_L} \cos(2\overline{w}s).
\end{equation}
Therefore, the Raman modulation to the intensity $\braket{I_{y'}}$ is the sum of the the following  contributions
\begin{equation}
i\int_0^{\tau} \mathrm{d}s B^y \braket{\Big[ H_{Ram}^{(1)}(s), a^\dag_{x k} a_{x k} \Big] } = i c_L |\alpha_k| \Big(|\alpha_{k-\frac{\Omega_E}{\delta}}|- |\alpha_{k+\frac{\Omega_E}{\delta}}| \Big) \braket{b^\dag - b}_{E_L} \sin^2(\tau (\overline{w}-w)) \frac{\sin(2\tau\overline{w})}{2\overline{w}},
\end{equation}
\begin{align}
&i\int_0^{\tau} \mathrm{d}s \Big( D^y \braket{\Big[ H_{Ram}^{(1)}(s), a^\dag_{x k} a_{y k} \Big] } + D^{y *} \braket{\Big[ H_{Ram}^{(1)}(s), a^\dag_{y k} a_{x k} \Big] } \Big) = \nonumber\\
&=- i c_L |\alpha_k| \Big(|\alpha_{k-\frac{\Omega_E}{\delta}}|- |\alpha_{k+\frac{\Omega_E}{\delta}}| \Big) \braket{b^\dag - b}_{E_L} \frac{\sin^2(\overline{w}\tau)}{\overline{w}} \frac{\sin(2\tau(\overline{w}-w))}{2}
\end{align}
Instead, when computing the modulation of the intensity $\braket{I_{x'}}$ the second contribution get a minus sign because $D^x=-D^y$ while the first one is obtained by substituting $B^y$ with $B^x=1-B^y$, namely $\sin^2(\tau (\overline{w}-w))$ with $\cos^2(\tau (\overline{w}-w))$. We can use the approximation $w\tau$ small as for LRM in order to get a better understanding of the different contributions. In this scenario, we see that the leading term is of order two for the modulation of $\braket{I_{y'}}$ while it is of order zero for $\braket{I_{x'}}$.

A similar analysis shows that for the transverse mode $E_T$ one has
\begin{equation}
\braket{\Big[ H_{Ram}^{(1)}(s), a^\dag_{x k} a_{y k} \Big] } = c_T |\alpha_k| \Big(|\alpha_{k+\frac{\Omega_E}{\delta}}| \braket{b^\dag}_{E_T} + |\alpha_{k-\frac{\Omega_E}{\delta}}| \braket{b}_{E_T}  \Big) ,
\end{equation}
\begin{equation}
\braket{\Big[ H_{Ram}^{(1)}(s), a^\dag_{y k} a_{x k} \Big] } =- c_T |\alpha_k| \Big(|\alpha_{k-\frac{\Omega_E}{\delta}}| \braket{b^\dag}_{E_T} + |\alpha_{k+\frac{\Omega_E}{\delta}}| \braket{b}_{E_T}  \Big)   ,
\end{equation}
\begin{equation}
\braket{\Big[ H_{Ram}^{(1)}(s), a^\dag_{x k} a_{x k} \Big] } = 0.
\end{equation}
As a consequence, the Raman modulation of $\braket{I_{y'}}$ is obtained as follows
\begin{align}
&i\int_0^{\tau} \mathrm{d}s \Big( D^y \braket{\Big[ H_{Ram}^{(1)}(s), a^\dag_{x k} a_{y k} \Big] } + D^{y *} \braket{\Big[ H_{Ram}^{(1)}(s), a^\dag_{y k} a_{x k} \Big] } \Big) = \nonumber\\
&= - c_T \tau |\alpha_k| \Big( |\alpha_{k+\frac{\Omega_E}{\delta}}| + |\alpha_{k-\frac{\Omega_E}{\delta}}| \Big)\braket{b^\dag + b}_{E_T} \frac{\sin(2\tau(\overline{w}-w))}{2} ,
\end{align}
while for the modulation of $\braket{I_{x'}}$ there is an opposite sign. The modulations of $\braket{I_{y'}}$ and $\braket{I_{x'}}$ are both of first order in $w\tau$.

\subsection{Time-dependence}
By specifying the structure of the matrix $\boldsymbol{\chi}^{(1)}$ we can compute explicitly the quantity $M(s)$ in \eqref{ms}. In particular, for the totalsymmetric mode $A$ we have
\begin{equation}
M(s)= a ,
\end{equation}
for the longitudinal mode $E_L$ we have
\begin{equation}
M(s)= c_L  \cos(2\theta) \cos(2\overline{w}s) ,
\end{equation}
while for the transverse mode the result is
\begin{equation}
M(s)= -c_T  \sin(2\theta).
\end{equation}
With all these ingrediends we can finally compute the averages $\braket{b}$ for the three modes. Explicitly they read
\begin{equation}
\braket{b}_A = -i a \tau \mathrm{e}^{-i \Omega_A t} \sum_j |\alpha^{pump}_{j}| |\alpha^{pump}_{j+\frac{\Omega_A}{\delta}}| ,
\end{equation}
\begin{equation}
\braket{b}_{E_L}= -i c_L \mathrm{e}^{-i \Omega_E t} \sum_j|\alpha^{pump}_{j}| |\alpha^{pump}_{j+\frac{\Omega_E}{\delta}}|  \cos(2\theta) \frac{\sin(2\overline{w}\tau)}{2\overline{w}},
\end{equation}
\begin{equation}
\braket{b}_{E_T}= i c_T \tau  \sin(2\theta) \mathrm{e}^{-i \Omega_E t} \sum_j|\alpha^{pump}_{j}| |\alpha^{pump}_{j+\frac{\Omega_E}{\delta}}| .
\end{equation}
By taking only the leading order in $w\tau$ the quantities $\braket{b}_A $ and $\braket{b}_{E_T}$ are unaffected while the expression for the $E_L$ mode becomes
\begin{equation}
\braket{b}_{E_L}= -i c_L \tau  \cos(2\theta) \,\mathrm{e}^{-i \Omega_E t} \sum_j|\alpha^{pump}_{j}| |\alpha^{pump}_{j+\frac{\Omega_E}{\delta}}|  .
\end{equation}
By defining the parameters $\eta^{pump}_A$ and $\eta^{pump}_E$ as follows
\begin{equation}
\eta^{pump}_A = \sum_j|\alpha^{pump}_{j}| |\alpha^{pump}_{j+\frac{\Omega_A}{\delta}}| , \quad \eta^{pump}_E = \sum_j|\alpha^{pump}_{j}| |\alpha^{pump}_{j+\frac{\Omega_E}{\delta}}| 
\end{equation}
one can simply write the quantities to be substituted in the previous results 
\begin{align}
\braket{b + b^\dag}_{A} &= - a \tau \eta^{pump}_A  2\sin(\Omega_A t), \quad &i \braket{b^\dag-b}_{A}= - a \tau \eta^{pump}_A  2\cos(\Omega_A t), \\
\braket{b + b^\dag}_{E_L} &= -c_L \tau \cos(2\theta) \eta^{pump}_E   2\sin(\Omega_E t), \quad &i \braket{b^\dag-b}_{E_L}=- c_L \tau \cos(2\theta) \eta^{pump}_E   2\cos(\Omega_E t), \\
\braket{b + b^\dag}_{E_T} &= c_T \tau \sin(2\theta) \eta^{pump}_E   2\sin(\Omega_E t), \quad &i \braket{b^\dag-b}_{E_T}= c_T \tau \sin(2\theta) \eta^{pump}_E   2\cos(\Omega_E t).
\end{align}

\subsection{Comparison with the experiment}

In the end, the expressions to be compared with the experiment are the difference between the transmitted intensity of the probe at delay time $t$ and the transmitted intensity without pump $\braket{I_{\lambda' k}}_<$ (at negative delay-times). In other words, the equilibrium behaviour is subtracted in order to highlight the phonon-dependent modulation. For the cross polarization we finally get
\begin{align}
&\braket{I_{y' k}}(t)-\braket{I_{y' k}}_<=- c_T^2 \tau^2 |\alpha_k|\Big(2|\alpha_k| + |\alpha_{k+\frac{\Omega_E}{\delta}}| + |\alpha_{k-\frac{\Omega_E}{\delta}}| \Big)  w \tau\, (\cos(\phi)-1)\sin(2\theta) \eta^{pump}_E   2\sin(\Omega_E t) \\
\end{align}

Instead, for the observable $\braket{I_{x' k}}(t)-\braket{I_{x' k}}_<$ the leading terms are of zeroth order in $w\tau$ and are related to Raman scattering mediated by the modes $A$ and $E_L$, while first order modulations due to the $E_T$ mode are less relevant in this case
\begin{align}
&\braket{I_{x' k}}(t)-\braket{I_{x' k}}_<= -a^2 \tau^2 |\alpha_k| \Big(|\alpha_{k-\frac{\Omega_A}{\delta}}|- |\alpha_{k+\frac{\Omega_A}{\delta}}| \Big) \eta^{pump}_A  2\cos(\Omega_A t) +  \\
& -c_L^2 \tau^2 |\alpha_k| \Big(|\alpha_{k-\frac{\Omega_E}{\delta}}|- |\alpha_{k+\frac{\Omega_E}{\delta}}| \Big) \cos(2\theta) \eta^{pump}_E  2\cos(\Omega_E t) \\
& -c_T^2 \tau^2 |\alpha_k|\Big(2|\alpha_k| + |\alpha_{k+\frac{\Omega_E}{\delta}}| + |\alpha_{k-\frac{\Omega_E}{\delta}}| \Big)  w \tau\, (1-\cos(\phi))\sin(2\theta) \eta^{pump}_E   2\sin(\Omega_E t) .
\end{align}
These results can be used to discuss the experimental plots in the main text that we report here for convenience.

In particular, the plots in Figs.~\ref{Spectral_perp} and ~\ref{Spectral_par} refer to the cases where 
\begin{enumerate}
\item
one measures the transmitted probe light with polarization along the $y$ axis and the pump photons are polarized at $45^\circ$. The theoretical model correctly predicts that the leading modulation is due to an $E$ symmetry mode and does not involve a redistribution of the spectal weight among different frequencies;
\item
pump and probe polarizations are parallel to the $x$ axis and one measures the $x$-polarized transmitted light.  Both $A$ and $E_L$ modes contribute in this case, as correctly predicted by the model, and one can detect the Raman spectral shift in this case.
\end{enumerate}

\begin{figure} [htbp]
\centering
\includegraphics[width=12cm]{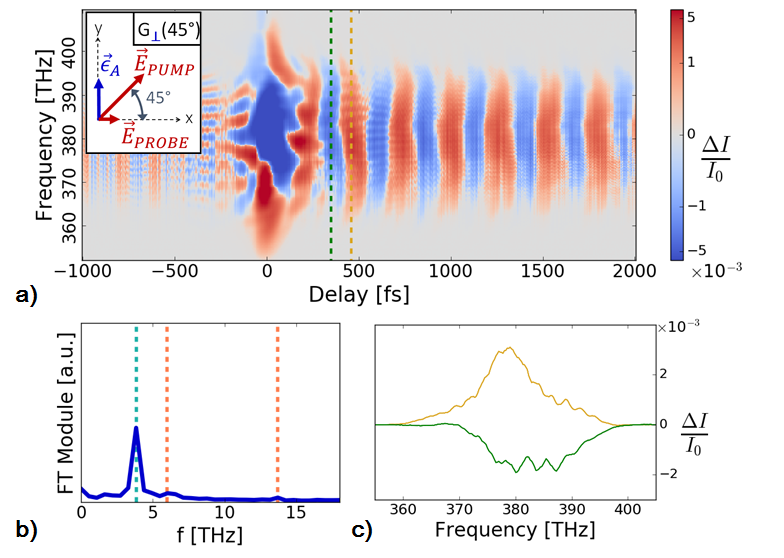}\\
\caption{a) Spectral modulation vs delay in the $(\lambda = y, \, \theta=45^\circ)$ configuration (insert) is presented. b) FT: only the 4 THz $E_T$ phonon is detected. c) The transmittivity modulation is selected at $t = 351$~fs (green) and $t = 458$~fs (gold). As expected for the refractive modulation the transmittivity change has the same sign for the whole spectrum. \label{Spectral_perp}}
\end{figure}

 \begin{figure} [htbp]
\centering
\includegraphics[width=12cm]{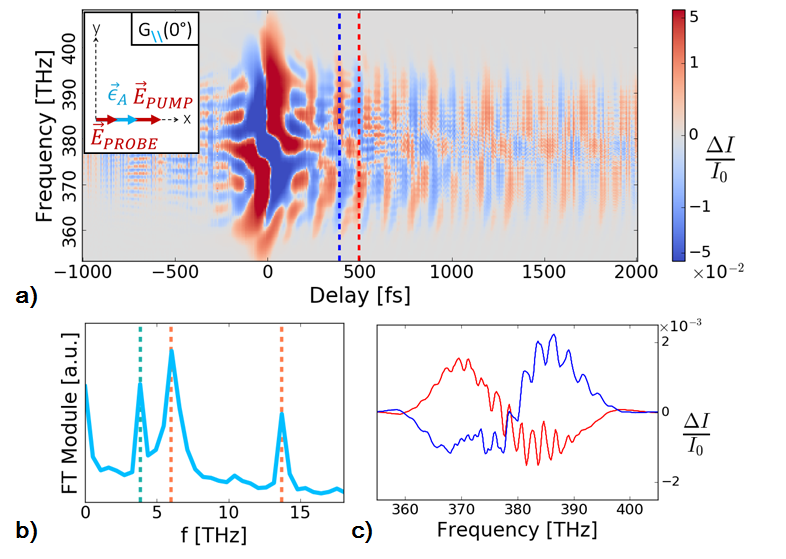}\\
\caption{Results depending on the relative orientation between pump and probe polarizations and analyzer direction. a) Spectral modulation vs delay in the $(\lambda = x, \, \theta=0^\circ)$ configuration (insert) is presented. b) FT: 4 THz $E_L$ phonon is detected, together with 6 THz and 14 THz $A$ symetry modes. c) The transmittivity modulation is selected at $t = 391$~fs (blue) and $t = 498$~fs (red). Spectral shifts resulting from ISRS are observed. \label{Spectral_par}}
\end{figure}

\section{Experimental details}

\begin{figure} [htbp]
\centering
\includegraphics[width=15cm]{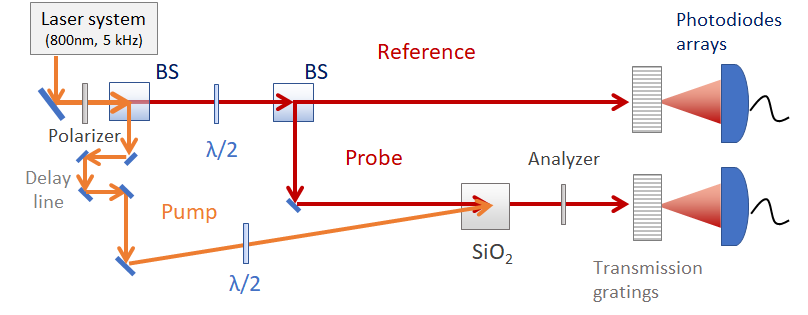}
\caption{Scheme of the experimental setup. \label{Setup}}
\end{figure}

In the main text, we report the results of pump\&probe frequency-resolved, polarization dependent, measurements on $\alpha$-quartz. In this section, a brief description of the experimental setup is given (Fig.\ref{Setup}).\\
The pulsed laser source is characterized by 800 nm wavelength, 5 kHz repetition rate and 40 fs pulse duration.\\
The output is split in order to obtain the pump and probe beams. The relative delay between the two is controlled by means of a translation stage on the pump path (delay resolution 6.7 fs). With a second beam splitter, a reference copy of the incident probe pulse is obtained, which is useful to remove noise and distinguish in the transmitted signal only the relevant information about the interaction processes with the sample.\\
Half-wave plates and polarizers control the relative orientation $\theta$ of pump-probe polarizations. A polarizer is also added after the sample (analyzer) to select the observed polarization $\lambda$.\\
The $\alpha$-quartz sample is 1 mm thick and the employed pump and probe fluences are respectively 0.8~mJ/cm$^2$ and 0.7~$\mu$J/cm$^2$.\\
Single-shot wavelength-resolved spectra of both probe beams are measured through a pair of transmission spectrometers, each provided with a linear array of 256 photodiodes, with frequency resolution 0.15 THz. The spectrum associated with a single time delay is the average over about 1000 single pulse acquisitions.

\end{document}